\documentclass[a4,useAMS,usenatbib,usegraphicx]{mn2e}
\def\aapr{\ref@jnl{A\&A~Rev.}}

\include{epsf}
\usepackage{float}
\usepackage{color}
\usepackage{latexsym}
\usepackage{amsmath}
\usepackage{amssymb}
\usepackage{multirow}
\usepackage{rotate}
\usepackage{epstopdf}
\DeclareGraphicsExtensions{.eps}

\epsfclipon

\title[Orbit \& Harassment]{The Sensitivity of Harassment to Orbit: Mass Loss from Early-Type Dwarfs in Galaxy Clusters}
\author[R.Smith et al.]{R.Smith$^{1,2,3}$\thanks{E-mail:rorysmith274@gmail.com},
R.S\'anchez-Janssen$^{4}$, 
M. A. Beasley$^{5,11}$,
G. N. Candlish$^{3}$,
B. K. Gibson$^{6}$,\and
T. H. Puzia$^{7}$,
J. Janz$^{8}$,
A. Knebe$^{9,10}$,
J. A. L. Aguerri$^{5,11}$,
T. Lisker$^{12}$,
G. Hensler$^{13,14}$,\and
M. Fellhauer$^{3}$,
L. Ferrarese$^{4}$,
and
S. K. Yi$^{1}$\\
$^{1}$Yonsei University, Graduate School of Earth System Sciences-Astronomy-Atmospheric Sciences, Yonsei-ro 50, Seoul 120-749, Republic of Korea \\
$^{2}$CEA-Saclay, DSM, DAPNIA, Service d'Astrophysique, 91191 Gif-sur-Yvette, France\\
$^{3}$Departamento de Astronomia, Universidad de Concepcion, Casilla 160-C, Concepcion, Chile\\
$^{4}$ NRC Herzberg Institute of Astrophysics, 5071 West Saanich Road, Victoria, V9E2E7, Canada\\
$^{5}$ Instituto de Astrofisica de Canarias, E-38200 La Laguna, Tenerife, Spain\\
$^{6}$ E.A. Milne Centre for Astrophysics, Dept of Physics \& Mathematics, University of Hull, Hull, HU6 7RX, UK\\
$^{7}$ Institute of Astrophysics, Pontifica Universidad Catolica, 7820436 Macul, Santiago, Chile\\
$^{8}$ Centre for Astrophysics \& Supercomputing, Swinburne University, Hawthorn, VIC 3122, Australia\\
$^{9}$ Departamento de Fisica Teorica, Facultad de Ciencias, Modulo C-8, Universidad Autonoma de Madrid, 28049 Cantoblanco, Madrid, Spain\\
$^{10}$Astro-UAM, UAM, Unidad Asociada CSIC\\
$^{11}$Universidad de La Laguna, Dept. Astrofisica, E-38206 La Laguna, Tenerife, Spain\\
$^{12}$	Astronomisches Rechen-Institut, Zentrum f\"ur Astronomie der Universit\"at Heidelberg, M\"onchhofstra\ss e 12-14, 69120 Heidelberg, Germany \\
$^{13}$	University of Vienna, Department of Astrophysics, Turkenschanzstr, 17 1180 Vienna, Austria\\
$^{14}$ National Astronomical Observatory of Japan, 2–21–1 Osawa, Mitaka-shi, Tokyo 181–8588}

\begin{document}

\date{Accepted to MNRAS: 8th September 2015}

\pagerange{\pageref{firstpage}--\pageref{lastpage}} \pubyear{2014}

\maketitle

\label{firstpage}

\begin{abstract}
We conduct a comprehensive numerical study of the orbital dependence of harassment on early-type dwarfs consisting of 168 different orbits within a realistic, Virgo-like cluster, varying in eccentricity and pericentre distance. We find harassment is only effective at stripping stars or truncating their stellar disks for orbits that enter deep into the cluster core. Comparing to the orbital distribution in cosmological simulations, we find that the majority of the orbits (more than three quarters) result in no stellar mass loss. We also study the effects on the radial profiles of the globular cluster systems of early-type dwarfs. We find these are significantly altered only if harassment is very strong. This suggests that perhaps most early-type dwarfs in clusters such as Virgo have not suffered any tidal stripping of stars or globular clusters due to harassment, as these components are safely embedded deep within their dark matter halo.
We demonstrate that this result is actually consistent with an earlier study of harassment of dwarf galaxies, despite the apparent contradiction. Those few dwarf models that do suffer stellar stripping are found out to the virial radius of the cluster at redshift=0, which mixes them in with less strongly harassed galaxies. However when placed on phase-space diagrams, strongly harassed galaxies are found offset to lower velocities compared to weakly harassed galaxies. This remains true in a cosmological simulation, even when halos have a wide range of masses and concentrations. Thus phase-space diagrams may be a useful tool for determining the relative likelihood that galaxies have been strongly or weakly harassed.
\end{abstract}


\section{Introduction}
Early-type dwarf galaxies are easily the most common galaxies in clusters (\citealp{VCC}). It is often stated that their low mass, and small potential well, should make them highly sensitive to their environment (e.g. \citealp{Lisker2009}). Environmental processes that could influence dwarf galaxies in clusters include ram-pressure stripping and starvation (\citealp{Boselli2008}), and harassment (\citealp{Moore1998}) from high speed tidal encounters with other cluster galaxies and the overall cluster potential. However the significance of each mechanism for early-type dwarf galaxies has yet to be ascertained (see \citealp{Recchi2014} for a recent review of environmental effects). 

Studies of dwarf ellipticals in Virgo -- the nearest galaxy cluster -- using the SDSS optical imaging have revealed them to be an inhomogenous group of objects (\citealp{Lisker2006}; \citealp{Lisker2007}). Some are shaped like thick disks, and may contain prominent spiral arms or bars, or blue cores of central star-formation. The degree of flattening is a function of luminosity, and the presence (or lack) of a nucleus (\citealp{Lisker2009}). Observations have also revealed that most early-type dwarfs are not morphologically simple (\citealp{Aguerri2005}; \citealp{Janz2014}); only $\sim$one fifth are well described by a single Sersic function with or without a nucleus. Roughly one third are found to have bars, and another $\sim$fifth have lens features (\citealp{Janz2012}; \citealp{Janz2014}). Dynamically, many early-type dwarfs  are rotationally supported (\citealp{Toloba2009}). In fact rotating early-type dwarfs lie on the same Tully-Fisher relation as dwarf irregulars (\citealp{Zee2004}; \citealp{Toloba2011}). However some early-type dwarfs have very complex dynamics including kinematically decoupled cores (\citealp{Toloba2014b}; \citealp{Toloba2014a}). Simulations demonstrate that harassment will rarely produce such features (\citealp{Gonzalez2005}), supporting the idea that tidal encounters in a group environment may be required to reproduce the dynamics of those galaxies presenting these complex dynamics.

In this study we will focus on how harassment drives the evolution of dwarf galaxies in the galaxy cluster environment. Harassment is the combined effects of tidal forces from the overall cluster potential well, and repeated high speed tidal encounters with other cluster galaxies. The effect of the high speed tidal encounters can enhance mass loss and heating by $\sim$$30\%$ (\citealp{Gnedin2003a}; \citealp{Knebe2006}). Tidal forces can strip dark matter, stars, and gas into low surface brightness streams that trail along the orbit of the galaxy through the cluster (\citealp{Smith2010a}). Low surface brightness disk galaxies are especially sensitive to harassment and may have their stellar disks truncated and heated (\citealp{Aguerri2009}), converting them into dispersion supported objects that resemble early-type dwarfs (\citealp{Moore1998}). However high surface brightness giant disks are much more robust to harassment (\citealp{Moore1999}), losing only a small fraction of stars, and suffering some disk thickening (\citealp{Gnedin2003a}). Indeed a difference in the size-luminosity relation of cluster disks compared to field disks is observationally detected. However \cite{Maltby2010} find a size difference only in intermediate and low luminosity disks, whereas in Coma a size difference is also detected in massive disks (\citealp{Gutierrez2004}; \citealp{Aguerri2004}). 

\cite{Mastropietro2005} conducted simulations of harassment acting on stellar-only disky dwarfs in clusters, by placing high resolution live models in a cosmological simulation of a cluster. Many of these initially thin, cool disk galaxy models ended up with thick, dispersion supported stellar disks following harassment. 
A similar process of tidal stirring is also believed to have occurred to initially disky dwarfs in order to form the dwarf spheroidal population of the Milky Way (\citealp{Mayer2005}). However, the \cite{Mastropietro2005} simulations only explored a limited number of orbits which, unfortunately, do not constitute an unbiased subsample from the full distribution of orbits occurring during cluster assembly. As we will show later, this has important implications for assessing the impact of harassment in the evolution of low-mass satellites.

Harassment should also have an impact on the globular cluster systems (GCSs) of dwarfs in galaxy clusters. A population of luminous early-type dwarfs in clusters have very rich globular cluster systems (\citealp{Miller2007}; \citealp{Peng2008}). \cite{Ruben2012} argued that the richness and extent of their GCSs is evidence against their progenitors being brighter, more massive galaxies. They also showed that at fixed stellar mass these massive early-type dwarfs also have a considerably larger GC population than field dwarf irregulars, which also argues against a direct link between these two populations\,\footnote{Note that, of course, this applies to these populations at the present day. The progenitors of current dEs were of course star-forming galaxies at some point in their past history, but the fossil record provided by their GCSs suggests that the early episodes of star formation in dEs occurred under different physical conditions from those in current field dIrrs--in the sense that GC formation was favoured.}. One example is VCC~1087 which has $>$60 GCs (\citealp{Beasley2006}), and these are distributed out to ~5-6~$R_{\rm{eff}}$.  If the GCS is very extended compared to stars, then the GCs are expected to be even more sensitive to tidal stripping than the stars. Then tidal encounters can strip off GCs (\citealp{Romanowsky2012}), and potentially place the GCS briefly out of dynamical equilibrium, and/or modify their radial profile about a cluster galaxy. Interestingly, the GCSs of some early-type dwarfs (e.g. VCC~1087, VCC~1261, VCC~1528) appear highly rotationally supported (\citealp{Beasley2009}). In \cite{Smith2013a} we studied the impact of harassment on the dynamics of the GCSs of early-type dwarfs undergoing harassment using numerical simulations. We found that roughly $\sim$$85\%$ of the dark matter halos had to be stripped before we saw any removal of stars or globular clusters. This occurs because the stars and globular clusters are embedded deeply within the potential well of their galaxy's dark matter halo, and so are not tidally stripped until the dark matter halo is heavily truncated, and there is little dark matter left. Similar results are found for models of dwarfs spheroidals suffering tidal stripping in the Local Group (\citealp{Penarrubia2008}). New recipes could be developed, based on these results, for improving the modelling of stellar stripping from galaxies in Semi Analytical Models. \cite{Rys2014} claim that galaxies with higher dark-to-stellar mass ratio are preferentially found in the cluster outskirts, whereas central objects show ratios that are lower. \cite{Smith2013a} also studied whether the GCS dynamics were too disturbed by tidal encounters for dynamical masses, derived from their motions, to be useful. We found that the GCSs very quickly relax after a tidal encounter, and that unbound GCs quickly separate from the stellar body of the early-type dwarf. Therefore the GCS dynamics provide good dynamical mass measurements, only failing when the early-type dwarf galaxy is on the verge of complete destruction ($>$95$\%$ of the dark matter unbound).

However \cite{Smith2010a} found that the strength of the effects of harassment is highly dependent on the orbital parameters of the galaxy. This is in good agreement with fully cosmological simulations (\citealp{Mastropietro2005}). Eccentric orbits that pass near the cluster centre but have apocentre near the cluster virial radius were found to typically suffer weak effects from harassment. In comparison, a circular orbit near the cluster centre, that spends a much greater fraction of its time where the potential is most destructive, suffers strong mass loss from harassment. \cite{Bialas2015} also find that, in addition to orbit, disk inclination with respect to the orbital plane and disk size can also influence the amount of stellar mass lost. \cite{Smith2013a} found that most of the mass loss arises from the tides of the cluster potential, with a small additional amount of mass loss ($\sim30\%$) due to high speed tidal encounters with other cluster galaxies, in agreement with previous studies (\citealp{Gnedin2003a}; \citealp{Knebe2006}). Observationally differences can be seen in the properties of cluster dwarfs if they are divided into two bins; high and low line-of-sight velocities. Those in the low velocity bin are systematically more round in shape than those in the high velocity bins, for a sample of galaxies at small projected distance from the cluster centre (\citealp{Lisker2009}). \cite{Muriel2014} show that for late-type galaxies between 1 and 2 virial radii from the cluster, a low velocity sample showed systematically redder colour, higher surface brightness, and smaller size, at fixed stellar mass, compared to a high velocity subsample.

Clearly orbital parameters are highly important for dictating the strength of the cluster environmental effects. However in \cite{Smith2010a} and \cite{Smith2013a} we considered only a limited range of orbits (e.g. circular near the cluster centre, or plunging from the cluster outskirts). Thus the aim of this study is to carry out a comprehensive study of how orbital parameters control the strength of harassment. We therefore run simulations for 168 different orbits (each with their own unique orbital parameters) of an early-type dwarf galaxy model undergoing harassment in a simulated cluster environment. Our set up in described in \S2, our results are given in \S3, and we summarise and conclude in \S4.

\section{Setup}

\subsection{The code}
In this study we make use of {\sc{gf}} (\citealp{Williams2001}; \citealp{Williams1998}), which is a TREESPH algorithm that operates primarily using the techniques described in \cite{Hernquist1989}. The early-type dwarf galaxies we consider are gas-free and therefore not star-forming. We therefore do not include an SPH component or star formation recipe in our models, and so we operate the code purely as a gravitational code without considering hydrodynamics. {\sc{gf}} has been parallelised to operate simultaneously on multiple processors to decrease simulation run-times. The tree code allows for rapid calculation of gravitational accelerations. In all simulations, the gravitational softening length, $\epsilon$, is fixed for all particles at a value of 100 pc, which is large enough to avoid artificial clumping, and equal to the value used in the harassment simulations of \cite{Mastropietro2005}. Gravitational accelerations are evaluated to quadrupole order, using an opening angle $\theta_c=0.7$. A second order individual particle timestep scheme was utilised to improve efficiency following the methodology of \cite{Hernquist1989}. Each particle was assigned a time-step that is a power of two division of the simulation block timestep, with a minimum timestep of $\sim$0.5 yrs. Assignment of time-steps for collisionless particles is controlled by the criteria of \cite{Katz1991}. For details of code testing, please refer to \cite{Williams1998}. 

\subsection{New harassment model}

Following the approach of \cite{Smith2010a}, we model the dynamical and time-evolving potential of a galaxy cluster using analytical potentials (also see \citealp{Knebe2005}). An analytical gravitational potential is used for the main cluster, and each individual `harasser galaxy' (cluster galaxies that can interact with a model galaxy) has its own unique analytical potential. This approach has advantages - the spatial resolution of gravity from harasser galaxies is effectively infinite. Also placing high resolution model galaxies in cosmological simulations (e.g. \citealp{Moore1999}; \citealp{Mastropietro2005}) is numerically challenging. In comparison, computing accelerations from analytical potentials is very fast. There are also disadvantages, however, as harasser galaxies move on fixed tracks and cannot respond to the gravity of the live model galaxy. For high velocity encounters this is of negligible consequence, but we are unable to accurately model low speed tidal encounters using this approach and for this reason we deliberately select orbits resulting in high speed encounters only (see \S\ref{orbits} for more details). 

In this study we greatly improve on the harassment model of \cite{Smith2010a} and \cite{Smith2013a}. In the new harassment model the properties of the main cluster halo and harasser galaxy halos are dictated by a cosmological simulation of a cluster. We shall refer to the main cluster halo as the `cluster halo' hereafter, in order to distinguish it from other halos. For each snapshot of the cosmological simulation we measure the properties (e.g. virial mass and radius) of all halos at that instant. Then for each halo we construct an analytical potential that mimics each individual halo's tidal field. By applying the tidal field from all halos simultaneously we construct the total tidal field of the cluster. Unlike in \cite{Smith2010a} this approach allows for the growth in mass and structural evolution of the cluster with time as it accretes new galaxies. Furthermore, the properties of individual harasser galaxies also evolve in time. 

The original cosmological N-body simulation was performed with an adaptive mesh refinement code {\sc{MLAPM}} (\citealp{Knebe2001}), and is fully described in \cite{Warnick2006} and \cite{Warnick2008}. Each dark matter particle has a mass $\sim$$1.6\times10^8 h^{-1} $M$_\odot$, and the highest spatial resolution is $\sim$2~kpc. The cluster used is C3 from Table 1 of \cite{Warnick2006}. At z=0 the cluster has a virial mass of 1.1$\times$10$^{14} h^{-1} $M$_\odot$, and a virial radius of 973$ h^{-1}$kpc (a reasonable approximation to Virgo, see Urban et al. 2011, McLaughlin 1999). The cluster is followed for $\sim$7~Gyr (since redshift=0.8), during which time it approximately doubles in mass. A snapshot is produced once every $\sim$$0.16$~Gyr. This high time resolution enables us to follow the orbits of individual halos with high accuracy. In any snapshot halo properties are measured with the halo finder MHF (MLAPM's halo finder; \citealp{Gill2004a}), down to 20 particles per halo. Therefore the minimum resolved halo mass is $\sim3$$\times10^9 h^{-1} $M$_\odot$. There are initially a total of 402 halos including the cluster and harasser halos.

In our harassment model, each halo's potential well is described by a Navarro, Frenk and White (NFW) analytical potential (\citealp{Navarro1996}): 

\begin{equation}
\label{potNFW}
\Phi = -g_cGM_{200}\frac{ln(1+(r/r_s))}{r}
\end{equation}
\noindent where $g_c=1/[{ln(1+c)-c/(1+c)}]$, $c$ is the concentration  parameter that controls the shape of the profile, $r_{\rm{s}}$ is a characteristic radial scalelength, and $M_{200}$ is the virial mass.

The properties and position of each halo in a snapshot are taken directly from the cosmological simulation using the \cite{Gill2004a} halo tracker. Between snapshots we linearly extrapolate halo positions and properties. As snapshots are only $\sim$0.16~Gyr apart this leads to smooth halo motion as a function of time. This procedure allows us to calculate the potential and accelerations at any position, and at any moment over the $\sim$$7~$Gyr duration of the simulation. 

By linearly extrapolating the position of each halo, halo velocities may be artificially reduced, especially near pericentre where the motion of the halo is least linear. We attempt to quantify the reduction in the orbital velocities using the following approach. We first calculate the orbit of a tracer particle in a host halo with the same mass and properties as the main cluster halo in our harassment model. We choose an orbit with orbital eccentricity of 0.8, and with pericentre distance (normalised by the cluster virial radius) of 0.2. As we will show in Fig. \ref{Probplot}, this is a high probability orbit in cosmological simulations of clusters. Because we choose a high eccentricity, low pericentre distance orbit, the effect of linear extrapolation should be relatively strong. We then measure the reduction in the true orbital velocity that would occur if we had used linear extrapolation between snapshots separated by 0.16~Myr, as done in our harasssment model. We find the orbital velocity is reduced by no more than $4\%$, and this only occurs when a halo is very close to pericentre where it spends little time. Therefore we consider the linear extrapolation to have a negligible impact on our results.

\subsection{Galaxy models}
Our approach is to place live models of galaxies on orbits within the harassment model. Our galaxy models consist of 3 components; a NFW dark matter halo, a thick exponential distribution of stars, and a spheroidal distribution of globular cluster particles following a Hernquist profile. 

\subsubsection{The dark matter halo}
The dark matter halo of the galaxy model has an NFW density profile:
\begin{equation}
\rho(r) = \frac{\delta_{\mathrm{c}} \rho_{\mathrm{crit}}}{(\frac{r}{r_{\rm{s}}})(1+\frac{r}{r_{\rm{s}}})^2}
\label{NFWdensprof}
\end{equation}
\noindent where $\delta_{\mathrm{c}}$ is the characteristic density and $\rho_{\mathrm{crit}}$ is the critical density of the Universe. Given that the NFW model has a divergent total mass, we truncate the profile at $r_{\rm{200}} = r_{\rm{s}} c$ where c is the halo concentration.

Positions and velocities are assigned to the dark matter particles using the publically available algorithm {\it{mkhalo}} from the {\sc{nemo}} repository (\citealp{McMillan2007b}). Dark matter halos produced in this manner are evolved in isolation for 2.5 Gyr to test stability, and are found to be highly stable.

Our model galaxy has a dark matter halo mass of $10^{11}$M$_\odot$, consisting of 100,000 dark matter particles, with a concentration $c = 14$. The virial radius is $r_{\rm{200}} = 95$~kpc with respect to the field. The peak circular velocity of the halo is 88~km~s$^{-1}$ at a radius of 15~kpc. If the halo were placed at a radius of 200~kpc within the smooth background potential of the cluster, the ratio of its central density is $\sim$5 times that of the surrounding ambient medium.

\subsubsection{The stellar distribution}
The stellar distribution of the galaxy has a radially exponential form (\citealp{Freeman1970}):
\begin{equation}
\label{expdisk}
\Sigma(R) = \Sigma_0 {\rm{exp}} (R/R_{\rm{d}})
\end{equation}

\noindent
where $\Sigma$ is the surface density, $\Sigma_{\rm{0}}$ is central surface density, $R$ is radius within the stellar distribution, and $R_{\rm{d}}$ is the exponential scalelength.

The scalelength and mass of the stellar distribution is chosen to approximately match the observed properties of luminous Virgo dEs. These are characterised by close-to-exponential luminosity profiles (\citealp{Janz2008}). Our fiducial model has a total stellar mass of $3.0 \times 10^{9}$M$_\odot$ ($3\%$ of the halo mass; \citealp{Peng2008}), and is formed from 20,000 star particles. The stellar distribution has exponential scalelength $R_{\rm{d}}=1.5~$kpc, consistent with the scalelength of disk galaxies of this stellar mass from \cite{Fathi2010}. The effective radius is $\sim$2.6~kpc. The effective surface brightness of the model is $\sim$20~mag/arcsec$^2$ in the H-band, which is reasonable for early-type galaxies of this luminosity (\citealp{Janz2014}).

\begin{figure}
\begin{center}
\includegraphics[height=9.0cm,width=9.0cm]{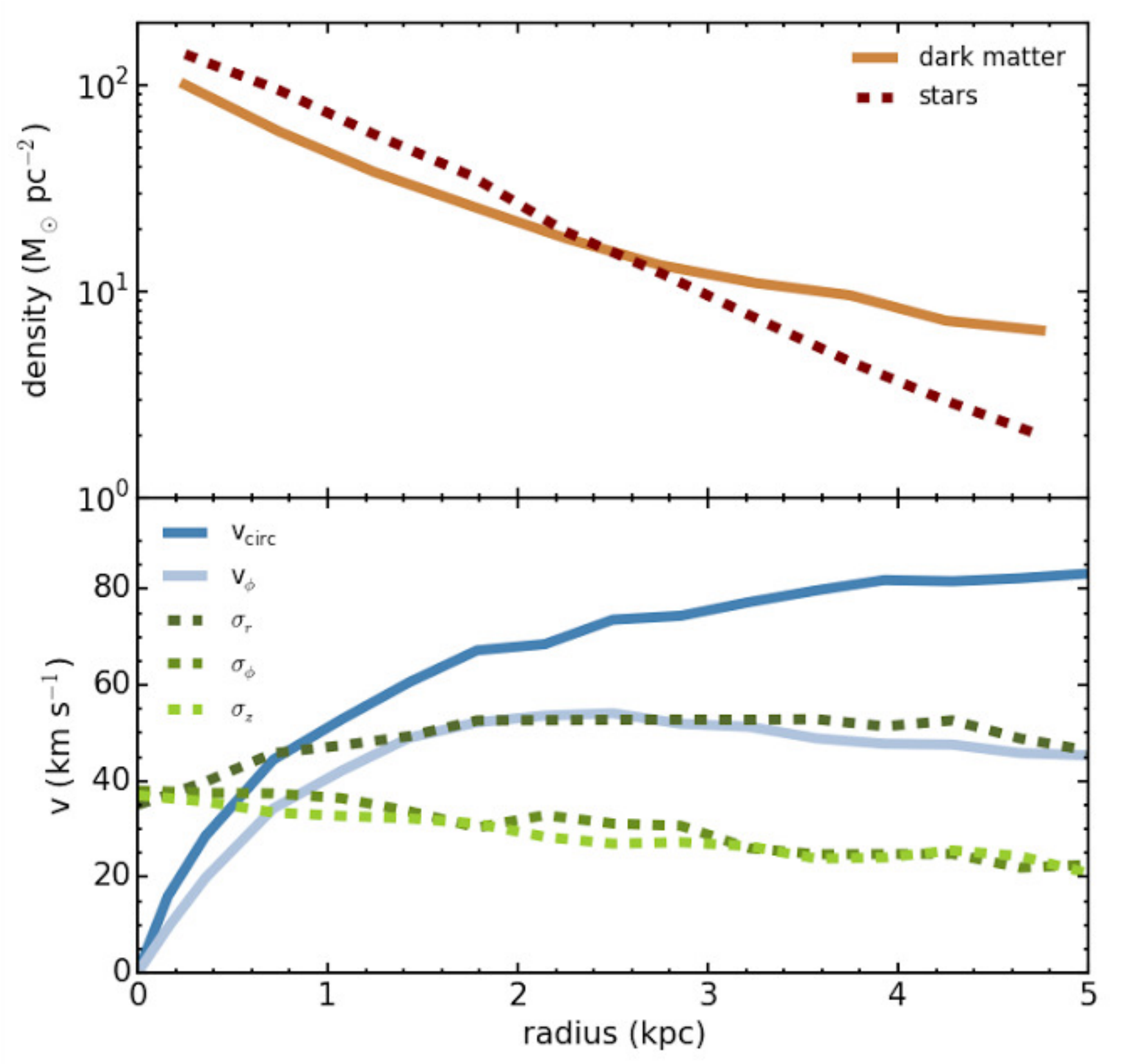}
\end{center}
\caption{The radial density profile of the dark matter halo (solid line), and stellar distribution (dashed line) is shown in the top panel. These quantities are measured within annuli that lie parallel to the plane of the disk, and whose vertical height extends to $\pm$1~kpc above and below the plane of the disk. In the lower panel, the circular velocity in the plane of the disk (solid, dark blue line) is shown, measured from the radial gradient of the potential. The azimuthal velocity (solid, light blue line), and the velocity dispersion components (dotted lines) are also shown, measured directly from the stellar dynamics.}
\label{radprofs}
\end{figure}

In the top panel of Fig. \ref{radprofs}, we show the radial density profile of the dark matter halo (solid line), and the stellar distribution (dashed line). In the bottom panel of Fig. \ref{radprofs}, the resulting circular velocity is shown (solid, dark blue line). This is similar to the circular velocity curves in observed early type dwarfs (\citealp{Rys2014}). We also show the azimuthal velocity (solid, light blue line), and the velocity dispersion components (dotted lines). A radially varying velocity dispersion is chosen that ensures the stellar distribution is Toomre stable (\citealp{Toomre1964}) at all radii. In practice, a Toomre parameter of $Q>1.5$ is required throughout the stellar body to ensure stability. We choose a fairly, hot stellar system with a similar degree of rotational and dispersion support in the plane. Early type dwarfs with a similar amount of dispersional support are observed in Virgo (\citealp{Toloba2009}; \citealp{Toloba2011}). This results in a thick stellar distribution with axial ratio b/a $\sim0.6$ consistent with the range of axial ratios observed in dwarf galaxies of this luminosity (\citealp{Lisker2007}; \citealp{Ruben2010}). A full description of the procedures followed to set up the stellar distribution can be found in \cite{Smith2010a}.

We choose a thick, hot stellar disk for our dwarf galaxy, unlike in many previous harassment studies whose stellar disks are initially cold, thin and highly rotationally supported (e.g. \citealp{Mastropietro2005}; \citealp{Smith2010a}; \citealp{Bialas2015}). The cool, thin disks of previous studies are often so thin that they have no known observable counterpart (\citealp{Ruben2010}). Harassment simulations that start with cool, thin dwarf disks demonstrate that to convert the thin disks into thick disks requires sufficent tidal heating that there is additional tidal stripping of the stars (e.g. see Figure 10 from \citealp{Mastropietro2005}). Our dwarf model approximates redshift zero observations of early type dwarfs with thick stellar disks, and rich extended globular cluster systems (e.g. VCC~1087). It is difficult to conceive how such dwarfs can initially have very thin disks, then lose enough mass that their stars are stripped and their disks thickened, without first stripping away their globular cluster system (cf. \citealp{Smith2013a}). Therefore it is likely that the types of galaxies we model formed with thick disks, or quickly evolved into thick disks through mechanisms other than heavy, harassment-induced tidal mass loss. Nevertheless, as we will discuss in \S\ref{mastrocompsect}, our choice of initially thick stellar disks has little significance for our key results on the amount of stellar stripping seen in our simulations. We will present a systematic study of stellar mass loss, and change in disk angular momentum, disk size, disk shape, and rotation-to-dispersion ratio, as a function of initial disk properties (e.g. disk thickness, disk size, halo profile, and initial rotation-to-dispersion ratio) in a follow-up publication.

\subsubsection{The globular cluster distribution}
We also include a component representing the globular cluster system of an early-type dwarf. Each globular cluster is treated as a single N-body particle. The standard model has 60 equal-mass globular cluster particles with a summed total mass of $1.8\times10^7$~M$_\odot$, corresponding to a specific stellar mass fraction of 0.6 per cent (\citealp{Peng2008}; \citealp{Georgiev2010}). We model the globular cluster particle distribution as a Hernquist sphere: 

\begin{equation}
\rho(r) = \frac{M_{\rm{h}}}{2\pi}\frac{r_{\rm{h}}}{r}\frac{1}{(r+r_{\rm{h}})^3}
\label{herneqn}
\end{equation}
where $M_{\rm{h}}$ is the total mass of the Hernquist sphere, $r_{\rm{h}}$ is the Hernquist scalelength, and $r$ is radius. 

The  distribution is radially truncated at a cut-off radius of $7.5~$kpc. The observed GCS density distribution in luminous dEs depends on projected radius roughly as R$^{-1}$ (\citealp{Beasley2006}; \citealp{Puzia2004}). We obtain an adequate match to this radial dependence by choosing the Hernquist scalelength $r_{\rm{h}}=3.75~$kpc (half the cut-off radius). Particle velocities are assigned using the Jeans equation for an isotropic dispersion supported system.

Finally, we note that all models (combined dark matter halo, stellar distribution and surrounding GCs) are evolved in isolation for 2.5~Gyr to ensure stability, before introduction into the cluster environment model.

\subsection{Orbits}
\label{orbits}
We choose to fully simulate only orbits that match the following requirements:

\begin{enumerate}
\item Objects on first infall are excluded -- galaxies must have had at least one pericentre passage.
\item Objects must have entered the virial radius of the cluster in the past.
\item Objects that entered the cluster in the past, but are now found out to twice the current virial radius are included -- backsplash galaxies (see \citealp{Gill2005}).
\item Objects that undergo low speed tidal encounters are excluded -- encounter velocities must be greater than 400~km~s$^{-1}$.
\end{enumerate}

In practice we find that galaxies that are on first infall do not result in any stripping of stars or globular clusters. In fact as we will show, even those that have passed pericentre once show no mass-loss of stars or globular clusters. Therefore we choose to exclude `first-infaller' orbits, as these consistently show no effects from harassment, and therefore we do not wish to spend computational time modelling systems where very little of interest occurs (requirement i). We also consider backsplash galaxies as these may have been affected by harassment when they were in the cluster, despite ending up outside of it (requirement iii). As discussed previously, harasser galaxies are on fixed orbital tracks and cannot respond to the potential of the live model. This is a reasonable approximation for a high speed encounter, but is a poor description of a low speed tidal encounter. Therefore we choose to model only the high speed tidal encounters of harassment, and exclude all orbits where low speed tidal encounters occur (requirement iv). With encounter velocities $>$400~km~s$^{-1}$, we ensure collisions occur at $>$5 times the internal velocities of the galaxy.

In order to find orbits that match these conditions, we place a massless tracer particle with the initial position and velocity of every halo in the cosmological simulation. We then evolve the orbits of all the massless tracer particles in the potential field of the harassment model. This step is necessary as the orbits of the massless tracer particles do not exactly match the orbits of the original dark matter halos, as the analytical potentials used in the harassment model are only approximations to the potential field of the original cosmological simulation. However by using massless tracer particles, we can quickly trace out all the orbits, and exclude those orbits that do not match our orbital criteria. We have confirmed that, when we replace a tracer particle with a live model of the galaxy, the tracer particle orbit accurately matches the orbit of the live model.
 
After filtering the orbits by these conditions, 168 out of 401 orbits remain. We simulate all of these with a live, high resolution galaxy model. For each orbit we measure the most recent apocentre and pericentre distance, and then use these to calculate the orbital eccentricity of the most recent orbit $e=(r_{\rm{apo}}-r_{\rm{peri}})/(r_{\rm{apo}}+r_{\rm{peri}})$. 

\subsection{Measured properties}
\subsubsection{Bound Fractions}
The following technique is found to be a reliable and reproducible method for measuring the fraction of particles that remain bound to the galaxy at any instant. Initially the position of the centre of density is obtained. This involves finding the particle with the highest number of nearby neighbours. In practice, for each particle, the number of neighbours within a sphere is counted. The radius of the sphere is systematically varied from 0.1-1.0~kpc in 9 equally spaced bins, and the position of the particle with the most neighbours is recorded for each of the 9 bins. An iterative one-sigma clip is then applied to these 9 positions, and the average of the positions that remain after the clipping is then used as the coordinates of the center-of-density. Then in the next step all particles within a 2.5~kpc radius of the centre-of-density of the galaxy are selected. The choice of 2.5~kpc is fairly arbitrary, but is roughly of order the effective radius of the dwarf galaxy and the final bound fractions are found to be very insensitive to this choice of radius. Each particle confirms whether it is bound to the other particles within this radius. Those that are bound to each other are considered a bound core within the galaxy center. Now an iterative procedure begins. In each iteration, all particles (including those beyond 2.5~kpc) are tested to see whether they are bound to the bound core. If they are, then their mass is added to the bound core. We call this method of growing the mass of the bound core the `snowballing' method. Then the iteration is repeated until the bound mass of the galaxy increases by no more than $1\%$ between iterations. In practice the total bound mass is found within 5-10 iterations. We find the snowballing method of measuring bound masses to be robust and trustworthy, and we have applied this technique successfully in several previous studies (\citealp{Smith2013c}; \citealp{Smith2013b}; \citealp{Smith2013a}).

\subsubsection{Generalised Sersic fits}
We measure the projected radial profiles of the stellar distribution and globular cluster distribution using a Sersic profile (\citealp{Caon1993}) for stars:

\begin{equation}
\Sigma(R)=\Sigma_{\rm{eff}} \exp \left(-b_{\rm{n}} \left[ \left(\frac{R}{R_{\rm{eff}}}\right)^{1/n}-1 \right] \right)
\label{genser}
\end{equation}
\noindent
where $n$ is the Sersic index, $b_{\rm{n}}=1.9992n-0.3271$, $R_{\rm{eff}}$ is the effective radius, and $\Sigma_{\rm{eff}}$ is the 
surface density at $R_{\rm{eff}}$. The globular cluster distribution is best modelled by an exponential distribution, similar to the observed globular cluster distributions of real dwarf galaxies (\citealp{Lotz2001}). Therefore we fix $n$=1 in order to measure the exponential scalelength of their distribution. Radial profiles are measured by placing a series of circular annuli in the plane of the disk, centred on the stellar density distribution of the model galaxy, and counting the mass in each annulus. Then a fit to the data is made using Eqn. \ref{genser}.

\begin{figure}
\begin{center}
\includegraphics[height=9.0cm,width=9.0cm]{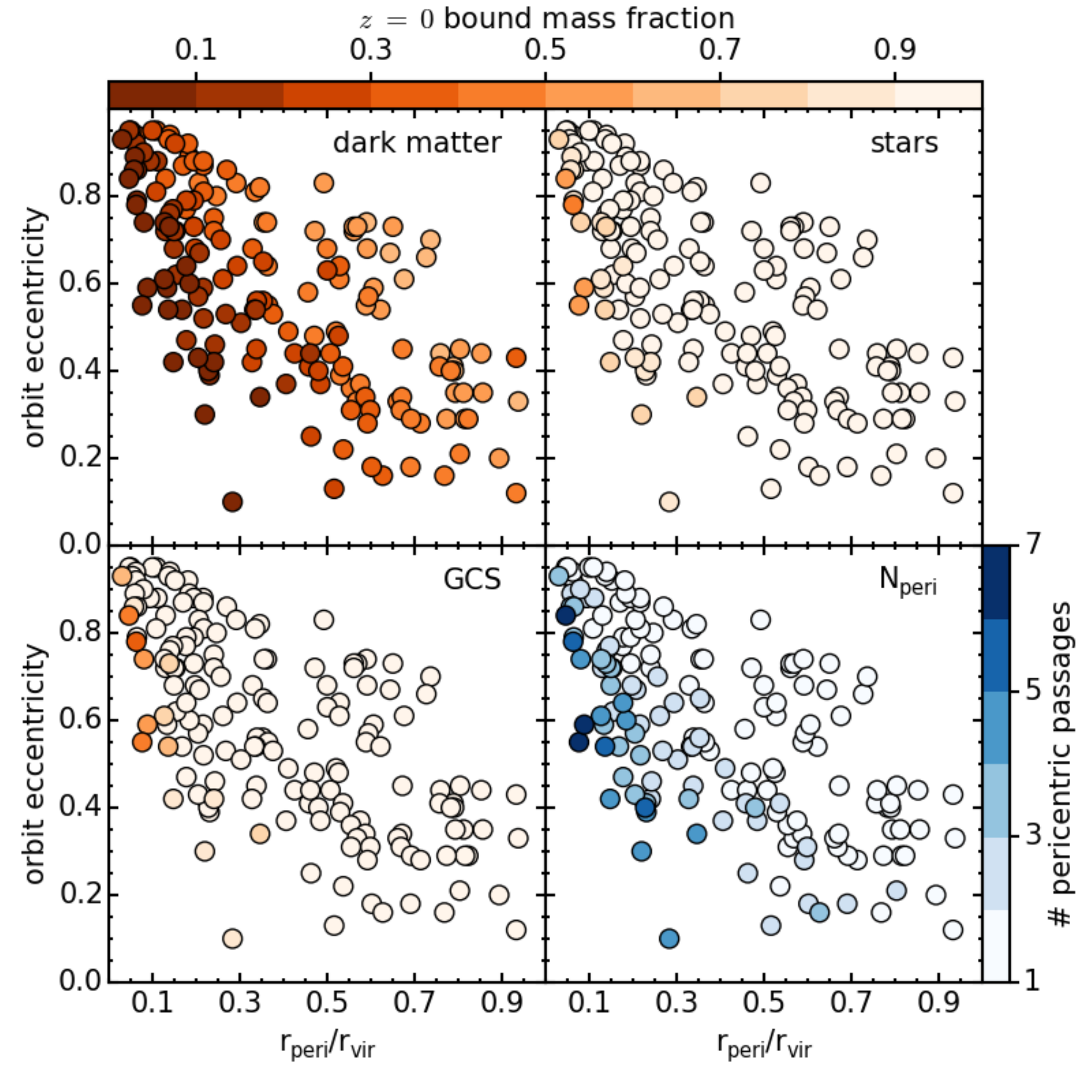}
\end{center}
\caption{The effect of orbital parameters on the bound mass fractions of all 168 of our models. Colour of the symbol indicates the final bound mass fraction of the model's dark matter (upper-left), stars (upper-right), and globular clusters (lower-left), using the upper colour-bar. In the lower-right panel the symbol colour indicates the number of peri-centre passages over the duration of the simulation (see lower-right colour bar). Each panel is a plot of the orbital parameters: eccentricity (y-axis) versus normalised pericentre distance (x-axis), where each symbol is one of our 168 galaxies.}
\label{orbitplots_masses}
\end{figure}

\section{Results}
\subsection{Dependence of mass-loss on orbital parameters}
In Fig. \ref{orbitplots_masses} we show the residual bound mass fractions at redshift z=0 (as symbol color shading) as a function of the orbital properties of each galaxy model (i.e. eccentricity on the y-axis, normalised pericentre distance on the x-axis) for the dark matter (upper-left panel), stars (upper-right panel), and globular clusters (lower-left panel). In the lower-right panel, the symbol colour shading indicates the number of pericentre passages over the duration of each harassment simulation.

\subsubsection{Dependence of dark matter mass-loss on orbit}
This is shown in the upper-left panel of Fig. \ref{orbitplots_masses}. The range of symbol colour shading across this figure indicates that dark matter mass-loss depends strongly on orbital parameters. Small pericentre distances (to the left of the panel) tend to suffer greater dark matter mass-loss. These destructive orbits can, however, have a wide range of eccentricity. Hence orbits with strong harassment fall in a tall and thin triangle, in the lower-left corner of the plot. If we classify strongly harassed galaxies as those that lose more than 85$\%$ of their dark matter (see Section\,3.1.2), the triangular shaped region reaches across to roughly $r_\mathrm{peri}/r_\mathrm{vir}$=0.4, and up to eccentricities of almost 1. This wide range of eccentricity means that strongly harassed galaxies must have a large range of apocentres, indicating that such galaxies will not be found exclusively near the cluster centre. 

\begin{figure}
\begin{center}
\includegraphics[height=5.5cm,width=9.0cm]{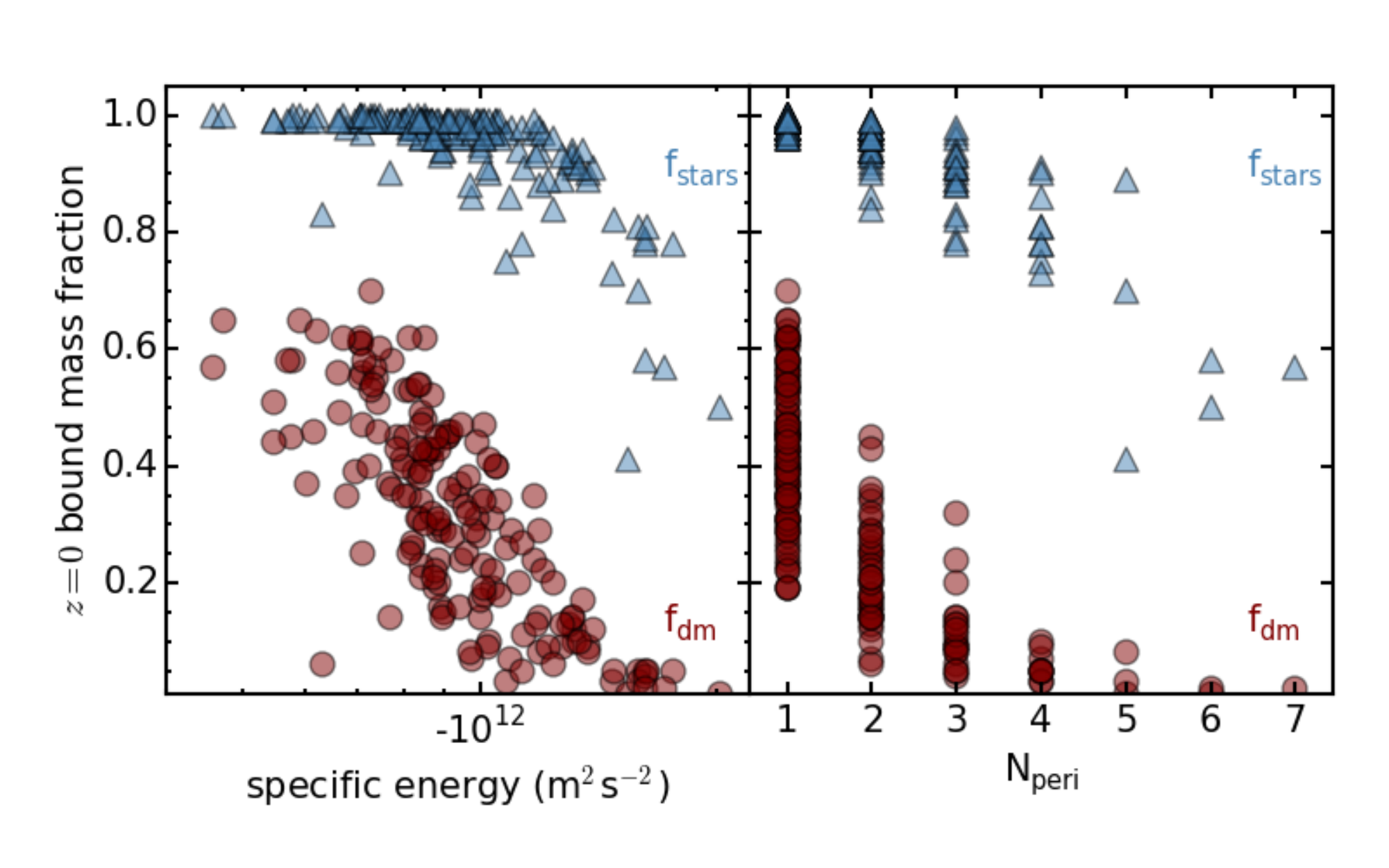}
\end{center}
\caption{Correlations between the final bound fraction of dark matter (red circles) and stars (blue triangles) as a function of orbital specific energy (left; shown as log-scale) and number of pericentre passages (right).}
\label{correlaciones}
\end{figure} 

To be as destructive as a circular orbit, a highly eccentric orbit must have a smaller pericentre distance -- this is the reason the region of strong harassment is triangular shaped. We will show in Sect.  \ref{neworbitdep} that the mass loss strongly depends on the pericentre distance and number of pericentre passages. There is a secondary dependence on eccentricity because, for a fixed pericentre distance, the eccentricity of the orbit will determine the number of pericentre passages. We note that here we have fixed the model galaxy's parameters and vary only the orbit. However in \cite{Smith2013a} we found neglible change to the bound dark matter fraction when we varied halo mass by a factor of 10, and a modest $\sim$15$\%$ change when we varied concentration from 5 up to 30.

\subsubsection{Dependence of stellar and GCS mass loss on orbit}
In the upper-right panel of Fig. \ref{orbitplots_masses} symbol colour indicates the final stellar bound mass fractions as a function of orbital parameters. Rather strikingly, {\it{the great majority of the orbits we consider result in no stellar mass-loss at all}}. Only models which suffer the very strongest dark matter mass-loss ($>$$85\%$ unbound), show any stellar mass-loss at all (those located in a tall, thin triangle in the lower-left of the distribution). This confirms our previous result -- that the stellar body only begins to be stripped once 80-90$\%$ of the dark matter halo has been unbound (e.g. see Fig.~5 of \citealp{Smith2013a}). This indicates that the stars are deeply embedded within the potential well of the halo, and so do not suffer tidal mass-loss until the halo has been heavily truncated and stripped itself.

In the lower-left panel of Fig. \ref{orbitplots_masses} symbol colour indicates the final globular cluster bound mass fractions as a function of orbital parameters. In many ways the bound GCS fraction panel is roughly the same as the bound stellar mass-loss panel (upper-right panel), demonstrating that GCs and the stellar distribution are similarly affected by harassment. This is because the stars and GCs have a similar radial distribution. As we will show in \S\ref{histmassloss}, however, the GCSs actually suffer marginally more tidal mass loss. Nevertheless, most of the orbits considered result in no loss of GCs. Only when the strongest dark matter mass-loss occurs (in the lower-left corner triangle described previously) are GCs stripped. Thus tidal stripping of stars and globular clusters requires orbits with small pericentre distances, but orbital eccentricity can vary widely. Furthermore, for very high eccentricity orbits the pericentre distance must be very small to cause tidal stripping of the stars and globular clusters.

\begin{figure}
\begin{center}
\includegraphics[height=5.5cm,width=9.0cm]{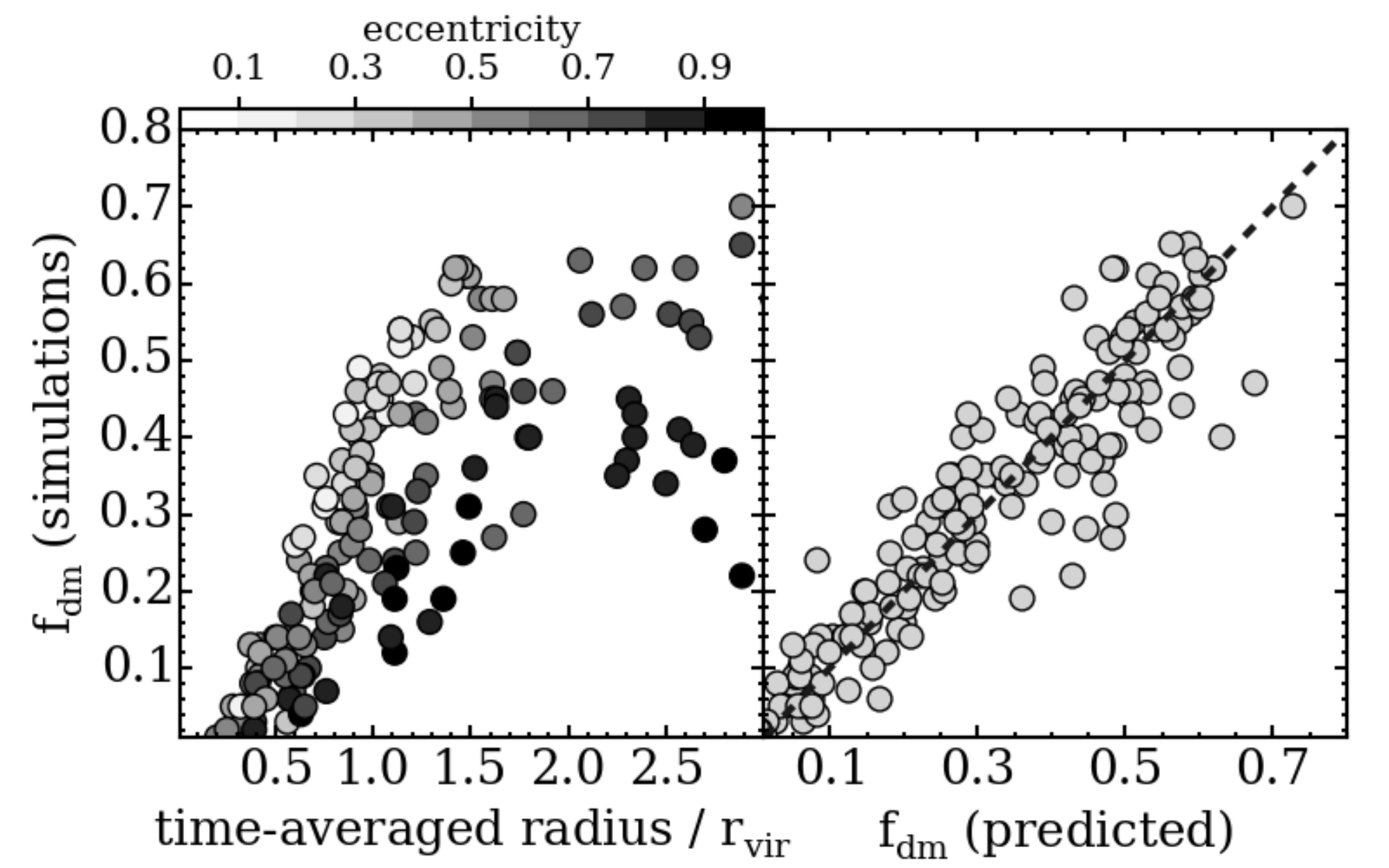}
\end{center}
\caption{Final bound dark matter fraction versus time averaged radius (left panel). Symbols are coloured by eccentricity (see colour bar). Final bound dark matter fraction versus vs predicted dark matter fraction (right panel) considering a simple mass loss model where there is a constant fractional mass loss between each pericentre passage (see text for details).}
\label{fdm_orbitalparams_correlation}
\end{figure} 

\begin{figure}
\begin{center}
\includegraphics[height=8.0cm,width=8.0cm]{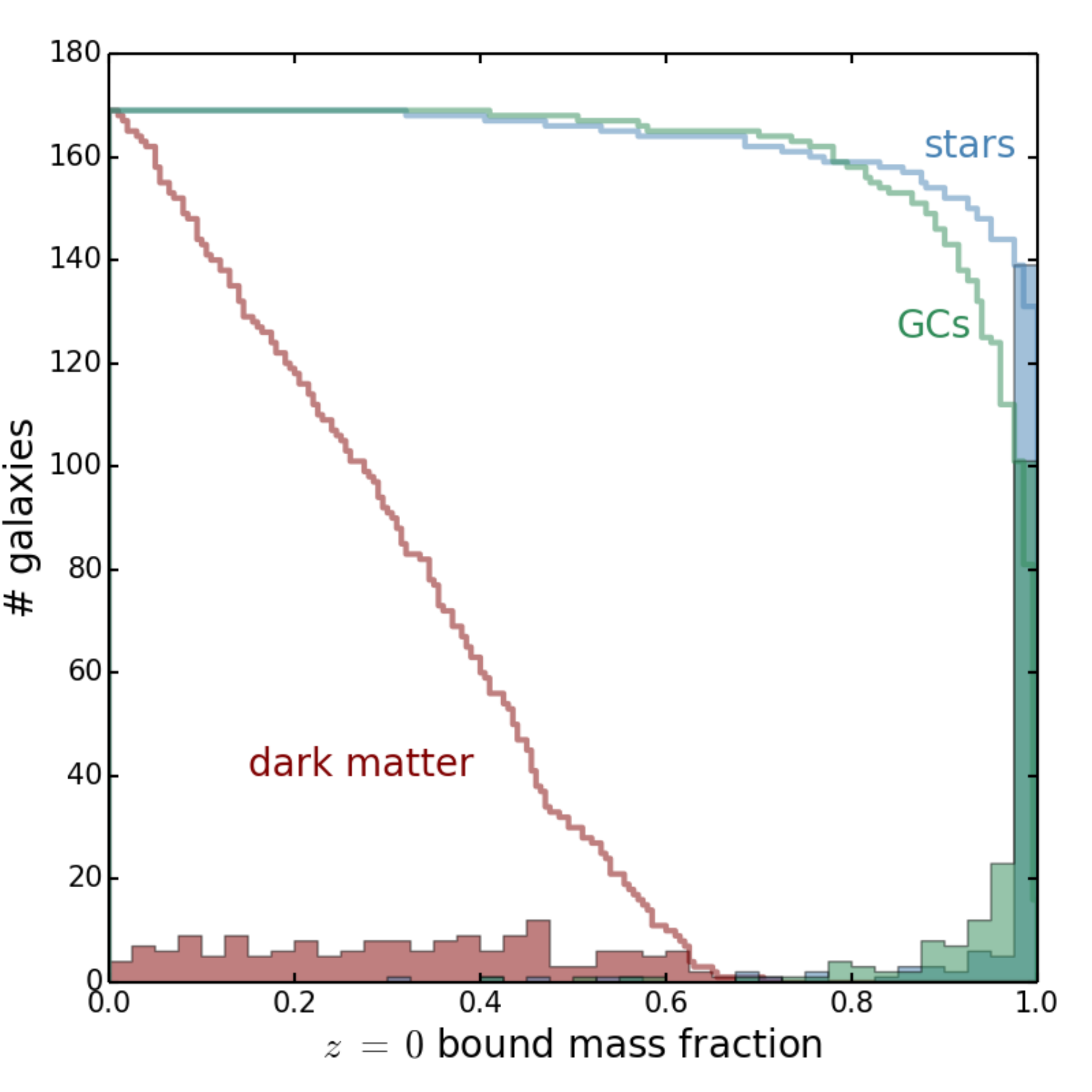}
\end{center}
\caption{The bound mass fractions at $z$=0 for all our model galaxies. Shaded histograms indicate the number distribution. Stepped lines indicate the cumulative distribution, which is cumulative moving from the right to the left. Colour indicates bound fractions of dark matter (brown), stars (blue), and globular clusters (green).}
\label{fmass_hist}
\end{figure}

\subsubsection{Mass loss as a function of specific energy and number of pericentre passages}
As shown in  the lower-right panel of Fig. \ref{orbitplots_masses} the great majority of our live models pass pericentre only once. This is because the cluster is constantly accreting new galaxies and so, at redshift zero, many galaxies may have only had time to be accreted and complete one pericentre passage. However where dark matter mass-loss was strong (in the lower-left corner triangle) galaxies suffer multiple pericentre passages (3 to 7 pericentre passages over the duration of the simulation). We further consider the correlation between number of pericentre passages and mass loss in the following subsections. \cite{Gill2004} find that with increasing numbers of pericentre passages, the orbits of substructure in their clusters become more circularised, due to the growth in the depth of the potential well of their clusters with time. This suggests that those of our models with large numbers of pericentre passages had higher eccentricity in the past.

In Fig. \ref{correlaciones} we consider the correlation between the final bound fraction of dark matter (red circles) and stars (blue triangles) as a function of orbital specific energy (left) and number of pericentre passages (right).  Orbital specific energy (left) is calculated by the sum of the kinetic and potential energy of each galaxy, measured at redshift zero. There is a strong correlation between the amount of dark matter that is stripped and the orbital specific energy. \cite{Rocha2012} find that specific orbital energy is also positively correlated with time of infall into the host halo. There is also a clear correlation between amount of dark matter and stars that are stripped and the number of pericentre passages the galaxy has completed by redshift zero (right panel). However, with each pericentre passage the fraction of the total bound mass that is stripped decreases rapidly. This is similar to the results of \cite{Taylor2004}, who finds that with each pericentre passage between a quarter and half of the {\it{remaining}} bound mass of a subhalo is stripped.

\subsubsection{Recipes for dark matter stripping as a function of orbital parameters}
\label{neworbitdep}
The time averaged radius of an eccentric orbit can be expressed as $r_{\rm{avg}}$$=$($r_{\rm{peri,norm}}/(1-e))(1-e^2)$, where $r_{\rm{peri,norm}}$ is pericentric distance normalised by the cluster virial radius, and $e$ is orbital eccentricity. In the left panel of Fig. \ref{fdm_orbitalparams_correlation} we plot final bound dark matter fraction versus time averaged radius. As the time averaged radius increases, the bound dark matter fraction increases. However the trend broadens at larger radius due to eccentricity, with high eccentricity galaxies have relatively lower dark matter fractions. Therefore the time averaged radius alone cannot be used as a reliable predictor for the amount of mass loss suffered.

In \cite{Taylor2004}, a simple recipe is considered for the mass loss of a satellite galaxy in a host halo. Between each pericentric passage, the mass of the satellite is found to decrease by between 25-45$\%$ of its mass prior to pericentric passage. The exact fraction that is lost, which we will refer to as $f_{\rm{peri}}$, is found to depend on the halo's concentration, and the eccentricity of the orbit. If these parameters are fixed, the fraction of dark matter remaining after $N_{\rm{peri}}$ pericentre passages will be: $f_{\rm{dm}}$=(1-$f$$_{\rm{peri}}$)$^{N_{\rm{peri}}}$. For our models, we know $f_{\rm{dm}}$ and $N_{\rm{peri}}$ (upper-left and lower-right panel of Fig. \ref{orbitplots_masses} respectively), so we can calculate $f_{\rm{peri}}$, in eccentricity and pericentre space. 

In practice, we find there is no clear trend in $f_{\rm{peri}}$ with eccentricity, when pericentric distance is fixed. The lack of a trend with eccentricity may arise because, with each pericentre passage, a galaxy's orbit becomes increasingly circularised by growth of the cluster potential (\citealp{Gill2004}). Instead, we find a clearer trend with pericentric distance. The trend has a linear form and can be approximated by $f_{\rm{peri}}=0.70 - 0.43$ $r_{\rm{peri,norm}}$. Therefore the mass lost at each pericentre can vary from $27\%$ (at $r_{\rm{peri,norm}}$=1.0) up to $70\%$ (at $r_{\rm{peri,norm}}$=0.0). Our upper limit of $f_{\rm{peri}}$=70$\%$ is higher than the 45$\%$ of \cite{Taylor2004}. However their recipe was for a satellite orbiting in a single host halo, and thus additional mass loss from harassment was not considered. 

Using this recipe, if we consider a galaxy with normalised pericentric distance $r_{\rm{peri,norm}}$, we can calculate $f_{\rm{peri}}$. Then, if the galaxy has $N_{\rm{peri}}$ pericentre passages, we can predict the final dark matter fraction. In the right-panel of Fig. \ref{fdm_orbitalparams_correlation}, we show the correlation between the predicted and the real dark matter fraction. A good correlation can be seen, and the predicted dark matter fractions matches the real dark matter fractions to within $\pm$0.11 (one-sigma errors).

In summary, a simple description for the distribution of dark matter mass loss in orbital parameter space can be found if we follow the approach of \cite{Taylor2004}. For a given orbit, we assume a constant fraction of the dark matter is lost between each subsequent pericentric passage. We find that this constant is a linear function of the pericentric distance of the orbit. In this description, with increasing numbers of pericentre passages, a galaxy suffers increasing amounts of mass loss. The number of pericentre passages is a function of both pericentre distance and eccentricity (e.g. see lower-right panel of Fig. \ref{orbitplots_masses}). In this way, the total mass loss becomes a function of both pericentre and eccentricity, although the fractional mass loss between pericentre passages is only a function of pericentre distance. With further testing, this recipe for dark matter stripping may be useful for semi-analytic models of galaxy formation where dark matter stripping is often prescribed based on simplistic simulations of dark halo interactions (e.g. \citealp{Lee2013}). We will test this prescription thoroughly in a future study, using cosmological simulations of clusters and groups with a wide range of masses.
 
\subsubsection{Histograms of mass loss}
\label{histmassloss}
In Fig. \ref{fmass_hist} we show the bound mass fractions by $z$=0 for all our sample of model galaxies. At z=0, all our halos have lost at least $\sim$30$\%$ of their dark matter (brown). There are roughly equal numbers in each bound mass bin between a bound mass fraction of 0.0 (all dark matter stripped), and 0.6 (only 40$\%$ of dark matter stripped). In comparison, the stars (blue) and globular clusters (green) suffer weaker fractional mass loss -- there are few models ($\sim$5$\%$) that lose more than $\sim$20$\%$ of their stars and globular clusters. This is because the stars and globular clusters are deeply embedded within the dark matter halo. Therefore the halo must be heavily truncated and stripped for the stars and globular clusters to suffer tidal stripping. In fact, it can be seen that the globular clusters are slightly more susceptible to tidal stripping than the stars. This is because in our initial conditions we assume the globular cluster distribution is slightly more radially extended than the stars. \cite{Wetzel2010} classify galaxies that have lost $\geqslant$97$\%$ of their dark matter as destroyed in their semi-analytical model prescription. Only 7 of our 168 models lose this much dark matter, and they lose 25-59$\%$ of their stars, and 28-68$\%$ of their GCS. However we note that the distribution of orbits found in our harassment simulations should not be considered identical to that found in cosmological simulations. Therefore to quantify the number of orbits that result in strong mass loss, we instead consider the fraction of such orbits found in cosmological simulations (see Sect. \ref{Howmany}).

\subsection{Final radial profiles of stars and the GCS - dependence on orbit}
\subsubsection{Effects on scalelengths of stars and the GCS}
\begin{figure}
\begin{center}
\includegraphics[height=5.5cm,width=9.0cm]{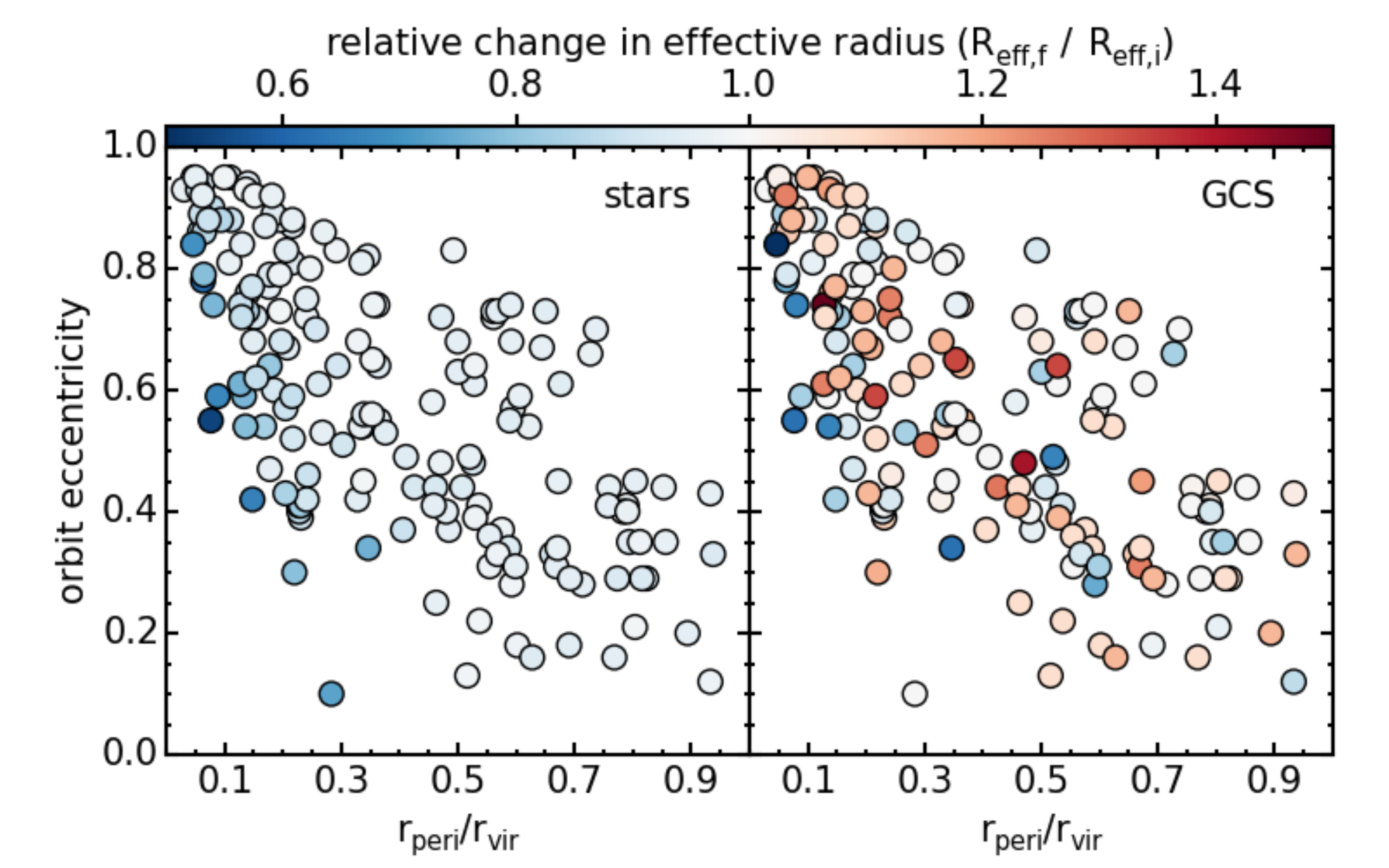}
\end{center}
\caption{Effect of orbital parameters on the scalelength of the spatial distribution of the stars (left panel) and globular clusters (right panel). Each panel is a plot of the eccentricity (y-axis) versus normalised pericentre distance (x-axis), where each symbol is a single galaxy from the harassment simulations. Symbol colour indicates the change in the effective radius.}
\label{orbitplots_sizes}
\end{figure}

In each panel of Fig. \ref{orbitplots_sizes} symbol position shows the orbital parameters of each model, as in Fig. \ref{orbitplots_masses}. We see that much of the orbital parameter space results in a stellar distribution that is entirely unaffected by harassment. Those stellar bodies that lose no stars, are also unchanged in effective radius. In the case of the strongest harassment, when stars are stripped, our models show reduced effective radii as their stellar distributions are truncated.

In the right panel symbol-colour indicated the exponential scalelength of the GCS distribution. Comparing to the stellar effective radius (left panel), the exponential scalelength of the GCS is less clearly correlated with orbit. There is a weak hint that the smallest scalelengths are found preferentially to the left of the distribution of points. However, in general the trend with orbit is very noisy due to the fact that, with only a few dozen globular clusters, we suffer low-number statistics when measuring the GCS scalelength. Also, by measuring variations of the inner GCS, we are probing changes in the innermost radii of the GCS, instead of the outer GCS which is more sensitive to harassment. We find clearer results by studying averaged number density profiles, as shown in \S\ref{GCprof}.

\begin{figure}
\begin{center}
\includegraphics[width=0.5\textwidth]{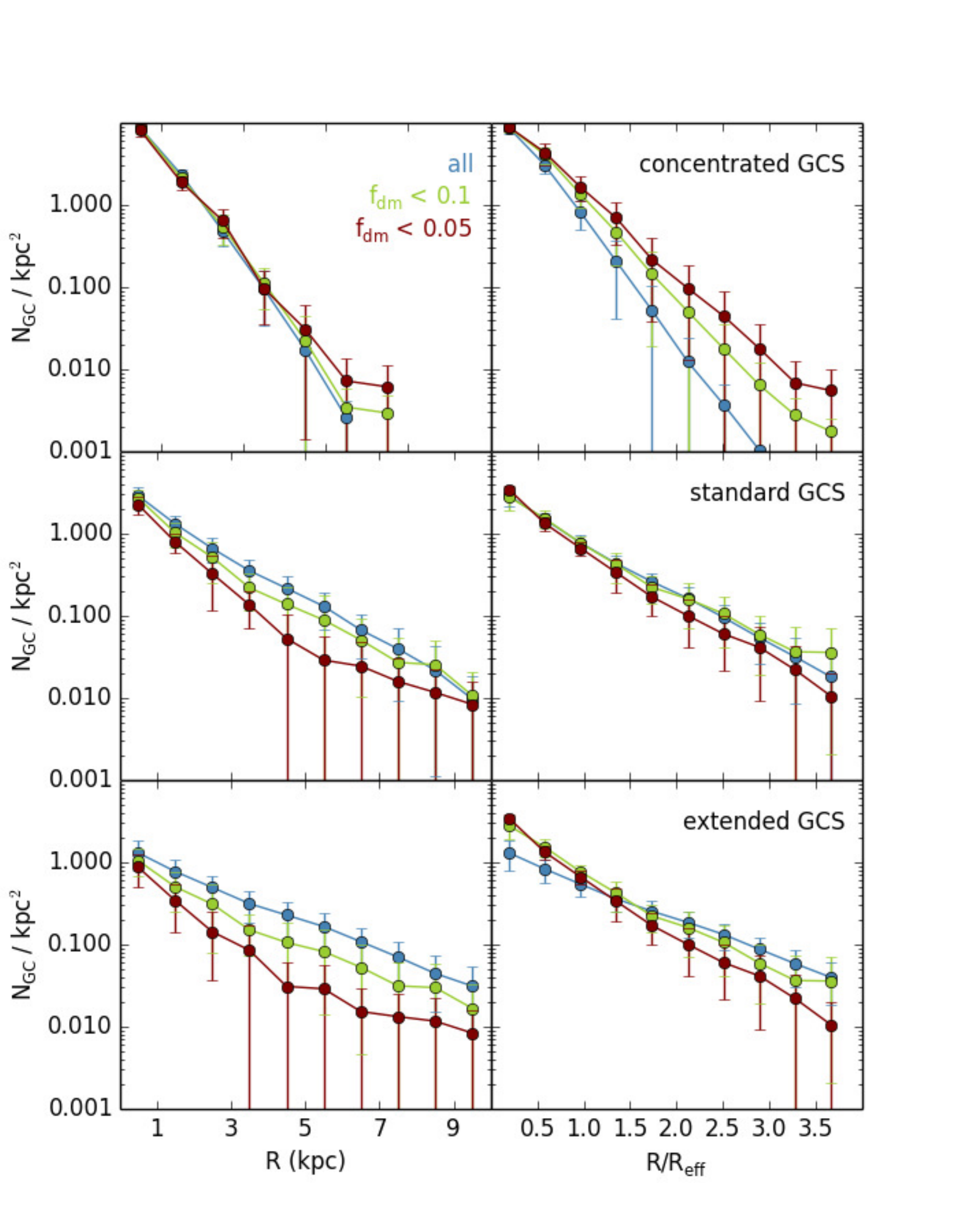}
\end{center}
\caption{Projected GC surface density profiles, averaged for all galaxies (blue curves), a sub-sample of strongly harassed galaxies (green curves) and very strongly harassed galaxies (red curves). Error bars indicate the standard deviation. The left column panels are the projected number density as a function of radius. The right-column panels show the projected number density as a function of galactocentric radius normalised by the effective radius of the stellar distribution. The distributions for three types of GCS concentrations are shown. The middle-row panels illustrate the results for the standard model used in this study, whereas the upper-row panels show the corresponding relations for an identical model except, in the initial conditions, the scalelength of the GCS distribution is halved creating a concentrated distribution. Similarly in the lower-row panels, the curves show the results when the scalelength of the standard model is doubled creating an extended GCS distribution.}
\label{GCSproffig}
\end{figure}

We measure a projected number density profile of the GCS for each model galaxy, by placing a series of circular annuli, centred on the centre-of-density of the stellar distribution. We then average the profiles for certain subsamples and illustrate the results in Fig. \ref{GCSproffig}. The blue lines are the whole sample, the green and red lines are for a sub-sample of strongly harassed ($f_{\rm{dm}}<$0.1) and very strongly harassed ($f_{\rm{dm}}<$0.05) galaxies respectively. Error bars indicate the standard deviation. Between the panel rows we vary how extended the GCS is initially. The middle row is our standard model with a Hernquist scalelength $r_{\rm{h}}$=3.75~kpc. In the upper row we halve the initial scalelength creating a concentrated GCS (`conc GCS') model, and in the lower row we double the initial scalelength creating an extended GCS (`extd GCS') model.

\subsubsection{Effect on the GC averaged number density profile}
\label{GCprof}

\begin{figure}
\begin{center}
\includegraphics[width=8.5cm]{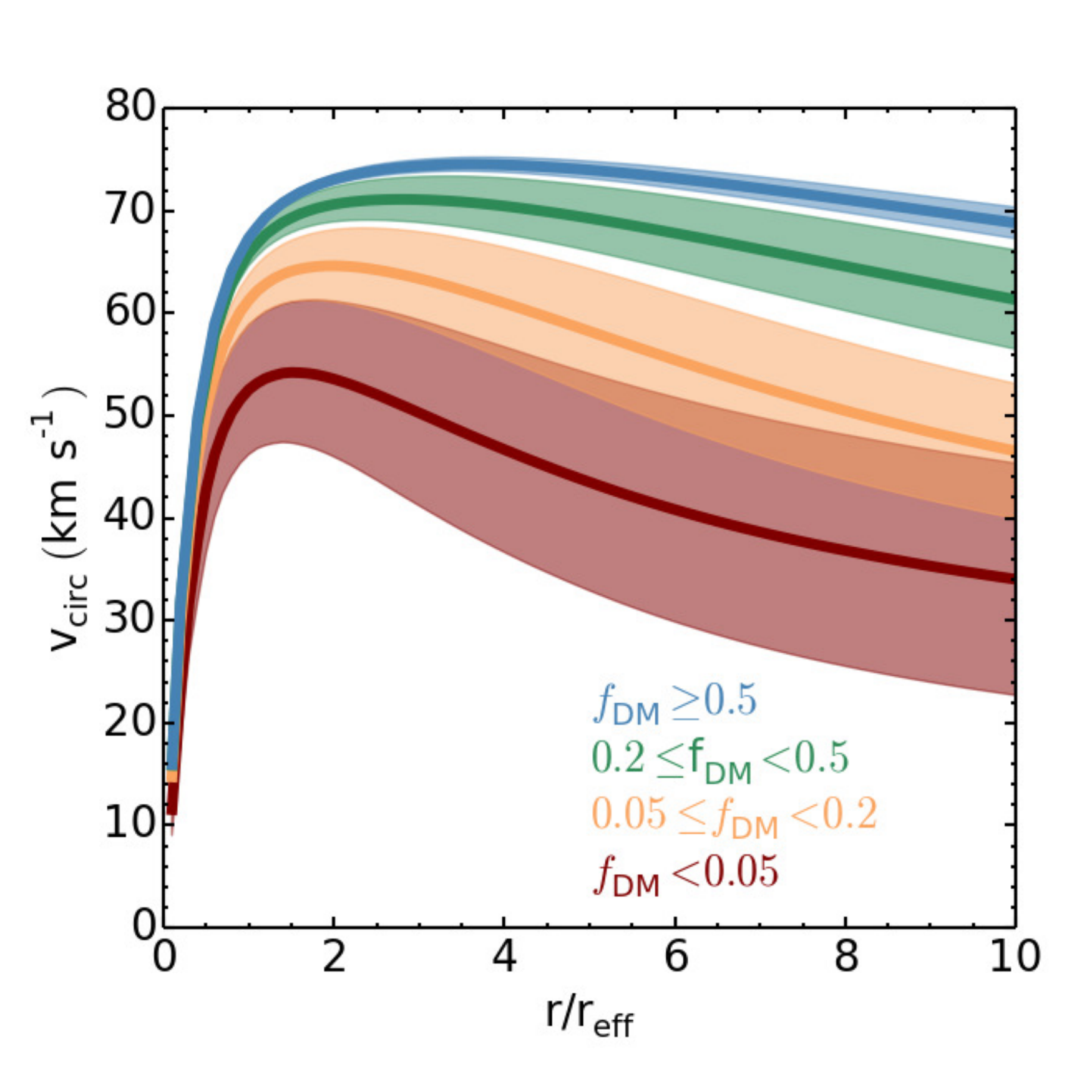}
\end{center}
\caption{The averaged circular velocity profile of our model dwarf galaxies, separated into subsamples according to the strength of dark matter mass loss that occured due to harassment (see legend). $f_{\rm{dm}}$$>$0.5 can be considered weak harassment with no stellar stripping (blue curve). $f_{\rm{dm}}$$<$0.05 can be considered very strong harassment resulting in significant stellar stripping (red curve). Shading indicates the one-sigma deviation of the subsample from the mean value.}
\label{vcirccurves}
\end{figure}

We first consider our standard model -- see the central-left panel. Even for strong harassment we find only a weak change in the GCS profile (compare the blue and green lines). However if harassment is {\it{very}} strong a significant change occurs, with the profile steepening at small radius, and flattening at large radius. Could this be used as a test of the strength of harassment in real galaxies? Unlike the model galaxy, real galaxies have a range of initial sizes. Therefore to create an average profile it is necessary to normalise the radius by the effective radius of the stars in the galaxy, as shown in the centre-right panel. The effective radius is measured at the same instant that the GCS is viewed. Unfortunately now the difference between the red and blue line is not so strong. This is because for very strong harassment, both the stellar disk and GCS are being affected by harassment in a similar way. When the GCS changes shape the effective radius is reduced, and this causes the change in profile shape to be less prominent when the normalised radius is used.

This is not the case if the stars and GCS are affected to differing degrees by harassment. For example, compare the middle-right panel to the upper-right and lower-right panels. If the GCS is initially very extended, it is preferentially truncated by harassment with respect to the stars, and then a change due to harassment is visible, even when the radius is normalised by the effective radius. Alternatively, if the GCS is initially very concentrated, then it is the stars that are preferentially truncated by harassment, while the GCS is nearly unaffected (see upper-left panel). Thus when the radius is normalised by the effective radius, the GCS appears to become {\it{more extended}} as a result of harassment.

The conclusion to draw from this is that the GCS profile is indeed altered when there is very strong harassment, at least in the standard and extended GCS case. However, when normalised radius plots are used, as is necessary for a real galaxy sample, detecting the change in profile becomes an additional function of how extended the GCS initially was with respect to the stars. {\it{If the GCS is initially very extended (considerably more than is currently observed) compared to the stars, then it is more sensitive to harassment than the stars. Then a change in GCS profile could be detected, in the case of very strong harassment.}}

\subsection{Changes to the circular velocity profile}

We calculate the total mass profile, measured in a spherical volume, as a function of radius for all of our galaxies at $z$=0. The total mass profile is converted into a circular velocity profile for each galaxy. We separate our galaxies into subsamples based on their final bound dark matter fraction, and combine the profiles together in each subsample to calculate an average circular velocity profile. The results are shown in Fig. \ref{vcirccurves}. The key indicates the range of bound dark matter fraction for each subsample. At large radius (r$\sim$10~$r_{\rm{eff}}$), the circular velocity is most sensitive to the mass loss, indicating preferential stripping of the outer halo first. Comparing the blue curve to the orange curve, we see that even when 80-95$\%$ of the dark matter is stripped (orange curve), the mean circular velocity has only fallen by $\sim$20$\%$ at 5~$r_{\rm{eff}}$. This indicates that the majority of the dark matter losses occurred beyond 5~$r_{\rm{eff}}$. However, if we consider the innermost circular velocity profile, there is very little change. This demonstrates that the innermost dark matter has been affected only weakly. A more significant change occurs to the circular velocity profile if harassment is very strong (i.e. resulting in more than 95$\%$ loss of dark matter (red curve). Now the mean velocity profile falls by $\sim$45$\%$ at 5~$r_{\rm{eff}}$, and the gradient of the profile at small radii begins to flatten. Although we note that, even with such strong harassment, the mean circular velocity in the inner 0.5~$r_{\rm{eff}}$ is still only slightly reduced.

\begin{figure}
\begin{center}
\includegraphics[width=8.5cm]{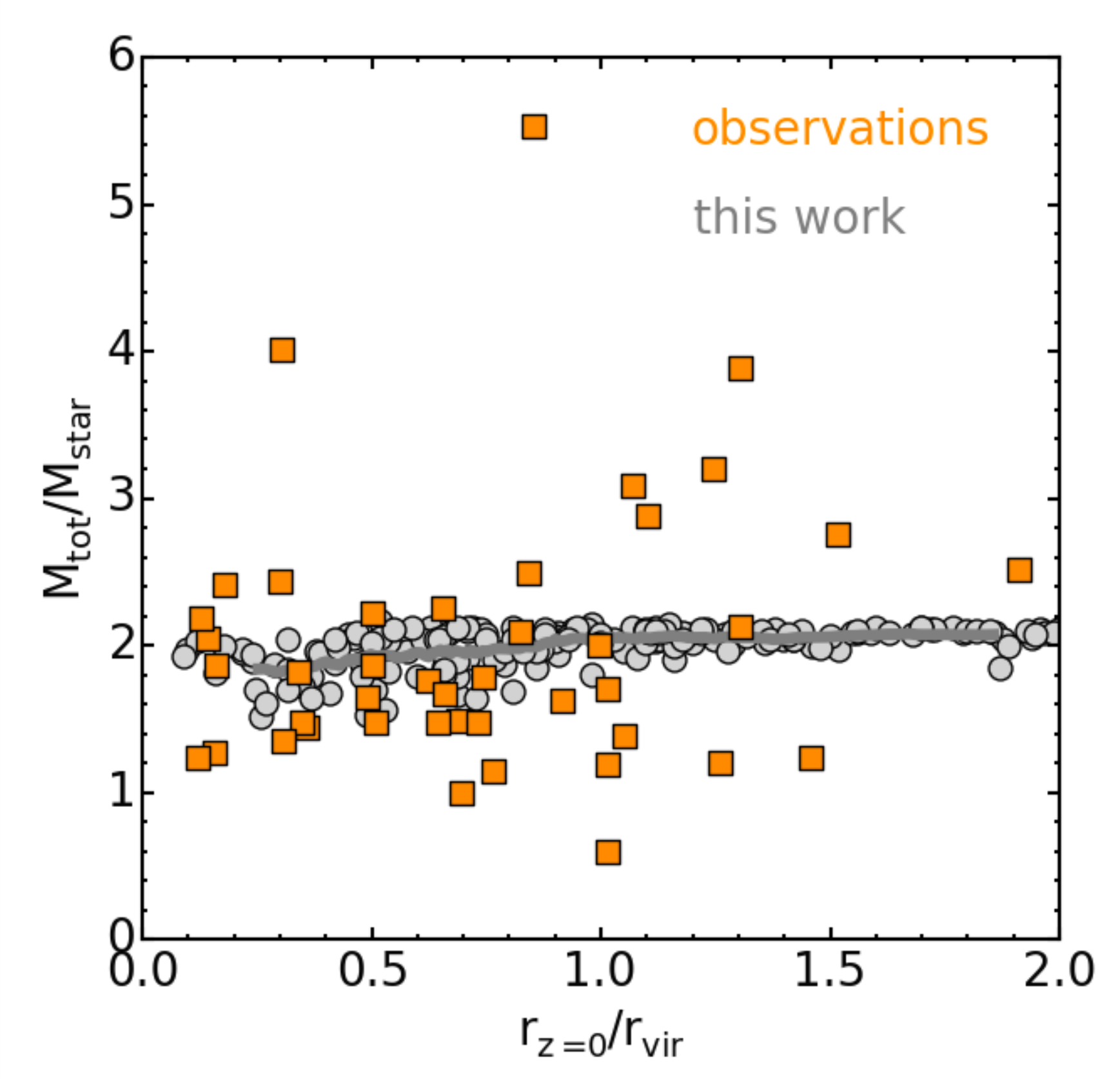}
\end{center}
\caption{The ratio of the total mass over the stellar mass, measured within one effective radius, is plotted against the normalised clustocentric radius for our model galaxies (grey symbols). All quantities are measured at $z$=0. The grey line indicates the running mean of our data points. Orange symbols indicate the observed results, compiled from several sources, by \citet{Penny2015}.}
\label{mass_frac}
\end{figure}

\subsection{Mass loss within one effective radius}
In Fig. \ref{mass_frac} we plot the ratio of the total mass over the stellar mass ($M_{\rm{tot}}/M_{\rm{\star}}$) within one effective radius as a function of a galaxy's normalised clustocentric radius. All quantities are measured at $z$=0 for all our galaxy models (grey symbols). At large radius (R$>$1.5~$R_{\rm{vir}}$) the ratio approaches $\sim$2.2, which is the value of our initial conditions. With decreasing radius the ratio falls in value to $M_{\rm{tot}}/M_{\rm{str}}$$\sim$1.6, with some scatter (values from $\sim$1.4-2.2). In general, our simulations show only a mild decrease in $M_{\rm{tot}}/M_{\rm{str}}$ as we move to smaller clustocentric radius. The gradient is shallow because mass is lost preferentially beyond one effective radius in our models (see previous section). These results can be compared with the data points of observed early-type dwarfs from Figure 8 of \cite{Penny2015}, given the caveat that the observed early-type dwarfs may not be subject to the constraint of having conducted at least one pericentre passage, as are our model dwarfs. We overlay the observed data points (orange symbols) on our plot. In general the trend between the observed galaxies and our simulations is similar. However there is clearly significantly less scatter in the simulation results compared to the observations. Although this is to be expected, given that we use exactly the same model dwarf galaxy for each harassment simulation. Also, our results are measured directly from the simulation and so do not have measurement errors. Meanwhile, the real dwarfs may have an intrinsic scatter in their properties, even prior to the effects of harassment. In addition, there is scatter due to measurement errors in $M_{\rm{tot}}/M_{\rm{str}}$. These measurement errors are typically greater than $\pm$0.5 (e.g. \citealp{Rys2014}; \citealp{Penny2015}).

\subsection{Searching for strongly harassed galaxies}
\subsubsection{How many are strongly harassed?}
\label{Howmany}
We have previously seen that `strong harassment' (i.e. harassment sufficiently strong to cause at least some stripping of stars and globular clusters), only occurs in a tall triangle in the lower-left corner of the eccentricity-pericentre distance plots (e.g. see upper-right panel of Fig. \ref{orbitplots_masses}). Therefore strong harassment occurs preferentially for orbits with small pericentre distance, but can have a range of eccentricity (hence the triangle extends up to eccentricity $\sim$1.0). To get an approximate estimate of the fraction of galaxies that fall within the `strong harassment' area of the orbital parameter space, we return to the 8 cosmological clusters described in \cite{Warnick2006}. All 8 clusters have similar, Virgo cluster-like masses ($\sim$(1-3)$\times$10$^{14}$~M$_\odot$). 

\begin{figure}
\begin{center}
\includegraphics[height=7.8cm]{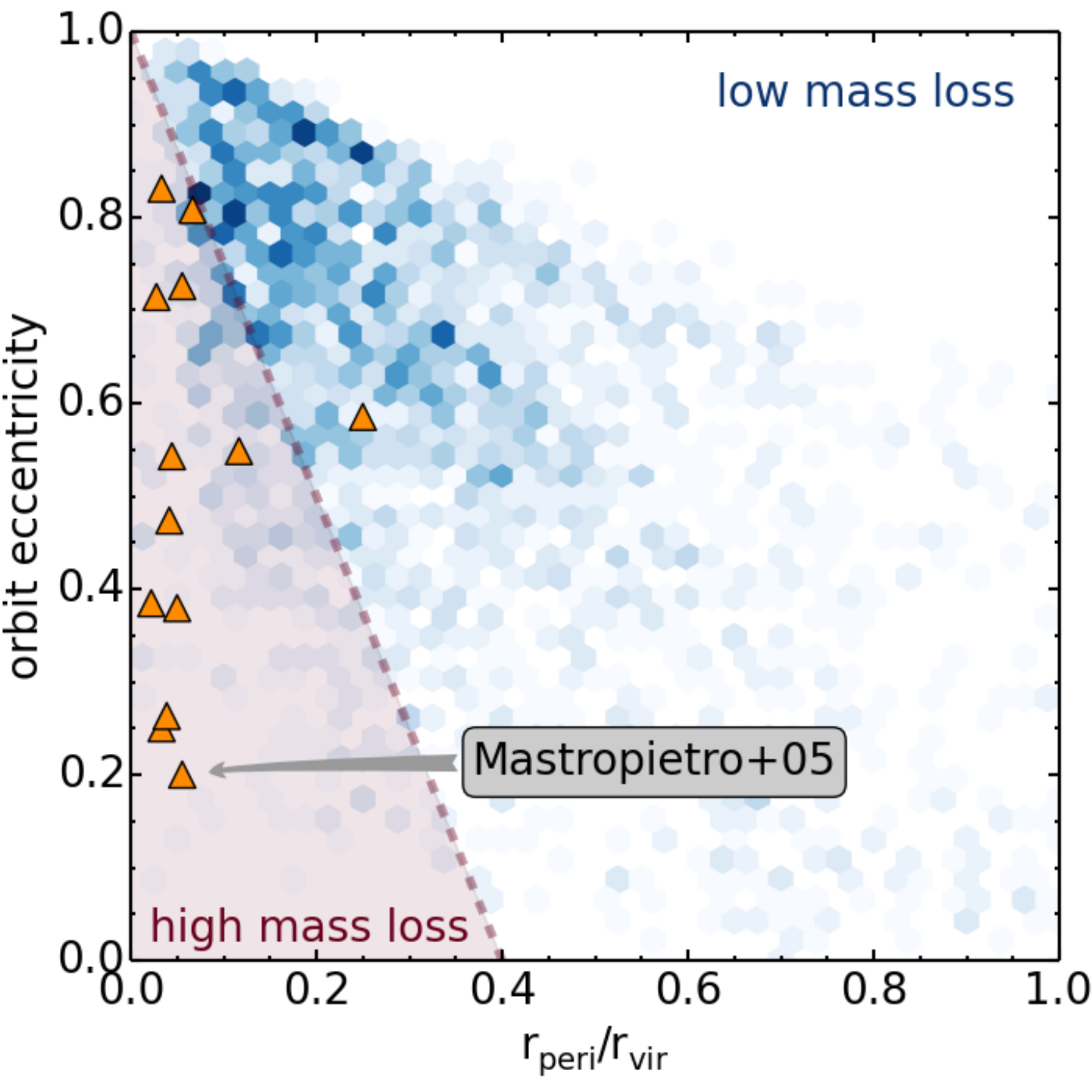}
\end{center}
\caption{Relative probabilities of different orbital parameters (eccentricity on the y-axis, normalised pericentre distance on the x-axis). Darker pixel indicates orbital parameters that are more commonly found in the 8 cosmological cluster simulations from \citet{Warnick2006}. All halos that were on first infall are excluded from this plot in order to make it comparable to these harassment simulations. The shaded region beneath the dashed line indicates the region of orbital parameter space where we see sufficiently strong harassment to begin stripping the stellar body of the galaxy in our harassment models (e.g. see upper-right panel of Fig. \ref{orbitplots_masses}). Orange triangles indicate the orbital parameters for the 13 galaxies from Mastropietro et al. (2005) that have experienced at least one full orbit.}
\label{Probplot}
\end{figure}

We first define a region in the orbital parameter space in which we see strong harassment occurring. This is simply done by eye, based on where we see stripping of stars, truncation of the stellar distribution, and unbinding of GCs (i.e. from the upper-right, lower-left panel of Fig. \ref{orbitplots_masses}, and left-panel of Fig. \ref{orbitplots_sizes}). Clearly doing this by-eye is rather approximate, but we only wish to get a rough estimate of the fractions of halos in cosmological simulations with these types of destructive orbits. We choose a line that intersects the y-axis at an eccentricity of 1.0, and intersects the x-axis at $r_{\rm{peri}}/r_{\rm{vir}}$=0.4, and every point below this line is assumed to lie in the strong harassment area. For every halo in the cosmological simulations we measure the eccentricity and pericenter distance of the most recent orbit.

Fig. \ref{Probplot} shows the distribution of orbital parameters from the cosmological clusters. We exclude `first-infaller' orbits in order to compare with our harassment models. The darkness of the blue shading indicates the relative numbers of halos that fall in any one pixel. The dark blue region indicates that orbits with eccentricities of $\sim$0.6-1.0 and pericentre distance of $\sim$0.0-0.2 are relatively more common than other orbits, in our cosmological clusters. This is in good agreement with the eccentricity and pericentre distributions in other cosmological simulations (\citealp{Benson2005}; \citealp{Wetzel2011}). The halo finder used (see \citealp{Gill2004a}) tracks halos even after they have been destroyed. Therefore the lack of orbits in the lower-left corner of Fig. \ref{Probplot} indicates that such orbits are genuinely rare. However the lack of orbits in the upper-right corner is due to the fact that we only consider orbits that conduct at least one pericentre passage, and there is not enough time for objects with such orbits to fall into the cluster.

We calculate the total percentage of halos that fall within the `high mass loss' area shown in Fig. \ref{Probplot}. If we consider halos that are found out to two virial radii at redshift zero (i.e. including the backsplash population) we find 19$\%$ lie in the triangle of orbits where at least some stars were stripped (i.e. the final bound stellar fraction is less than one). If we consider just the C3 cluster in the same way (which is the cluster we have used for the harassment model) the percentage is 18$\%$ indicating that cluster-cluster variations are not large. Once again we consider all the clusters together, but this time we include only halos found within one virial radius at redshift zero (i.e. excluding the back-splash galaxies). The percentage in the triangle is now 25$\%$, and therefore it is reasonable to state that less than a quarter of the orbits we measure in our cosmological simulations would result in stellar stripping.

In any case, the great majority of halos in our cosmological simulations do not have orbits which result in strong harassment. As all our clusters are roughly the mass of Virgo-like clusters, we suggest that this may be true for early-type dwarfs in Virgo too. In other words, {\it{the majority of early-type dwarfs in Virgo may have only suffered weak harassment (with no stellar stripping, or change in disk size), despite spending many gigayears in the cluster environment.}} We note that if we had included first infallers, the percentage of orbits resulting in strong harassment would likely be even lower, strengthening our conclusion even further. One caveat, however, is that we have assumed dwarf galaxies initially have thick, dispersion supported stellar disks. A second caveat is that we assume they enter the cluster environment as initially unperturbed systems, whose mass is dominated by an extended dark matter halo, in common with previous harassment studies, and consistent with the results from halo abundance matching (\citealp{Guo2010}).

\subsubsection{Comparison with Mastropietro et al. (2005)}
\label{mastrocompsect}
The small fraction of strongly harassed dwarf galaxies we find appears to be strongly at odds with the results of \cite{Mastropietro2005}, which claimed that harassment was highly efficient at stripping and heating stars from their model dwarf galaxies. In this section we attempt to understand the source of these apparent differences.

\begin{figure}
\begin{center}
\includegraphics[width=8.9cm]{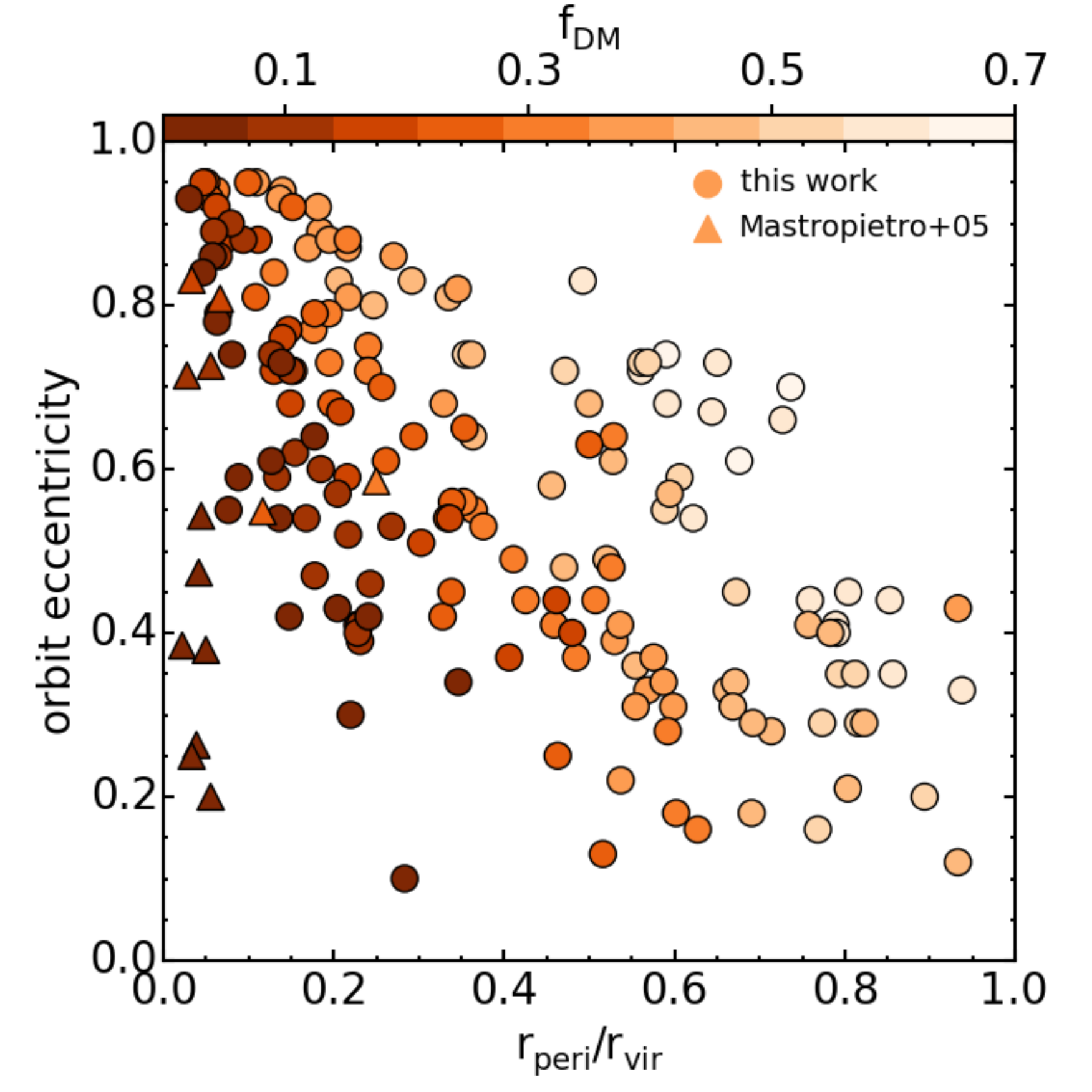}
\end{center}
\caption{Comparison of the dark matter bound fractions from our simulations (circle symbols) with the 13 galaxies available galaxies from Mastropietro et al. (2005; triangle symbols). Symbol shading indicates the remaining bound dark matter fraction, with darker shading showing lower remaining bound dark matter fractions (i.e. stronger mass loss).}
\label{mastrocomp_massloss}
\end{figure}

First note that, by construction, the 20 orbits simulated by \cite{Mastropietro2005} do not represent an unbiased subsample of the underlying orbital distribution for cluster haloes. On the contrary, half of their orbits correspond to particles that were already within the virial radius by $z=0.5$ and, more importantly, the other half were located within $0.2 \, r_{vir}$ at that time. As a result, their initial distribution of orbits is biased towards orbits that result in strong harassment.
This is shown by the orange triangles in Fig. \ref{Probplot}, where we plot all 13 galaxies from their simulations that have completed at least one orbit about the cluster by $z=0$ onto our probability density distribution of orbits (blue shading). It is clear that most of the galaxies have $r_\mathrm{peri}/r_\mathrm{vir}$$<$0.1 (with the exception of 2 galaxies). Furthermore, all but one fall within the triangle where we find harassment is sufficiently strong to affect the stellar disks of our model galaxies. Therefore it is perhaps unsurprising that for those 13 systems they find harassment to be highly influential for the stellar component of galaxies.
If we were to randomly select orbits from our cosmological simulations, it is highly unlikely that we would choose so many orbits falling within this triangle. This can be seen by comparing the location of the \cite{Mastropietro2005} data points to the distribution of orbits in our cosmological simulations. In fact, in the previous section we found that less than $\sim$20$\%$ of all subhalo orbits fall within this triangle, whereas this fraction is $\sim$60$\%$ in the \cite{Mastropietro2005} simulations. 

Second, one might wonder how well our results compare to Mastropietro's for galaxies on similar orbits about the cluster. In Fig. \ref{mastrocomp_massloss} the shading of each symbol indicates the amount of bound dark matter. Moving across the figure from lower-left to upper-right, bound dark matter fractions increase both in our simulations (circle symbols) and those from \cite{Mastropietro2005} (triangle symbols), in a similar manner. For similar types of orbits, the amount of dark matter remaining in our models is comparable with the amount of dark matter remaining in the \cite{Mastropietro2005} simulations. Thus our model dwarf galaxies are responding in a similar manner to the \cite{Mastropietro2005} dwarf models when they have similar orbits.

One clear difference between our initial conditions and those of \cite{Mastropietro2005} is that we choose a stellar disk that is thick, hot, and dispersion dominated, whereas their galaxy stellar disks are initially thin, cold, and rotation dominated. However this does not cause a significant change in the amount of tidal stripping of stars. For example, their thin cool disks do not lose significant numbers of stars ($>$10$\%$) until large amounts of dark matter ($>$85$\%$) have been stripped. This is almost identical to what is seen in our models (e.g. see Figure 5 from \citealp{Smith2013a}). Therefore disk thickness and the amount of dispersion support are apparently not sensitive parameters controlling stellar mass loss in early type dwarf disks. In addition \cite{Kazantzidis2011} find that tidal transformation of disky dwarf galaxies into dSphs occurs essentially independent of disk thickness.

Finally, it is important to point out that the seven (out of 20) galaxies in  \cite{Mastropietro2005} that have not completed one full orbit about the cluster by $z=0$ barely lose any stellar mass, and their disks suffer very little thickening or morphological transformation (other than the development of a bar, see Table 1 and Figure 10 in that paper). Unfortunately, these objects receive little attention throughout the subsequent analysis and discussion, giving the impression that the majority of cluster galaxies are strongly influenced by harassment.

In summary, we show that our model dwarfs suffer similar amounts of mass loss of dark matter as those considered in \cite{Mastropietro2005} {\it{if they have similar orbits}}. However, harassment appears much more influential in \cite{Mastropietro2005} as the type of orbits they consider are biased towards orbits that result in strong mass loss. Furthermore the majority of their models that suffer no stellar mass loss are neglected from the main body of their analysis. We find that changes in orbit can be very significant for the strength of harassment. This highlights the importance of considering the statistical probability of particular orbits when interpreting the results of harassment simulations, especially when a limited number of orbit types are considered.

\begin{figure}
\begin{center}
\includegraphics[height=9.cm]{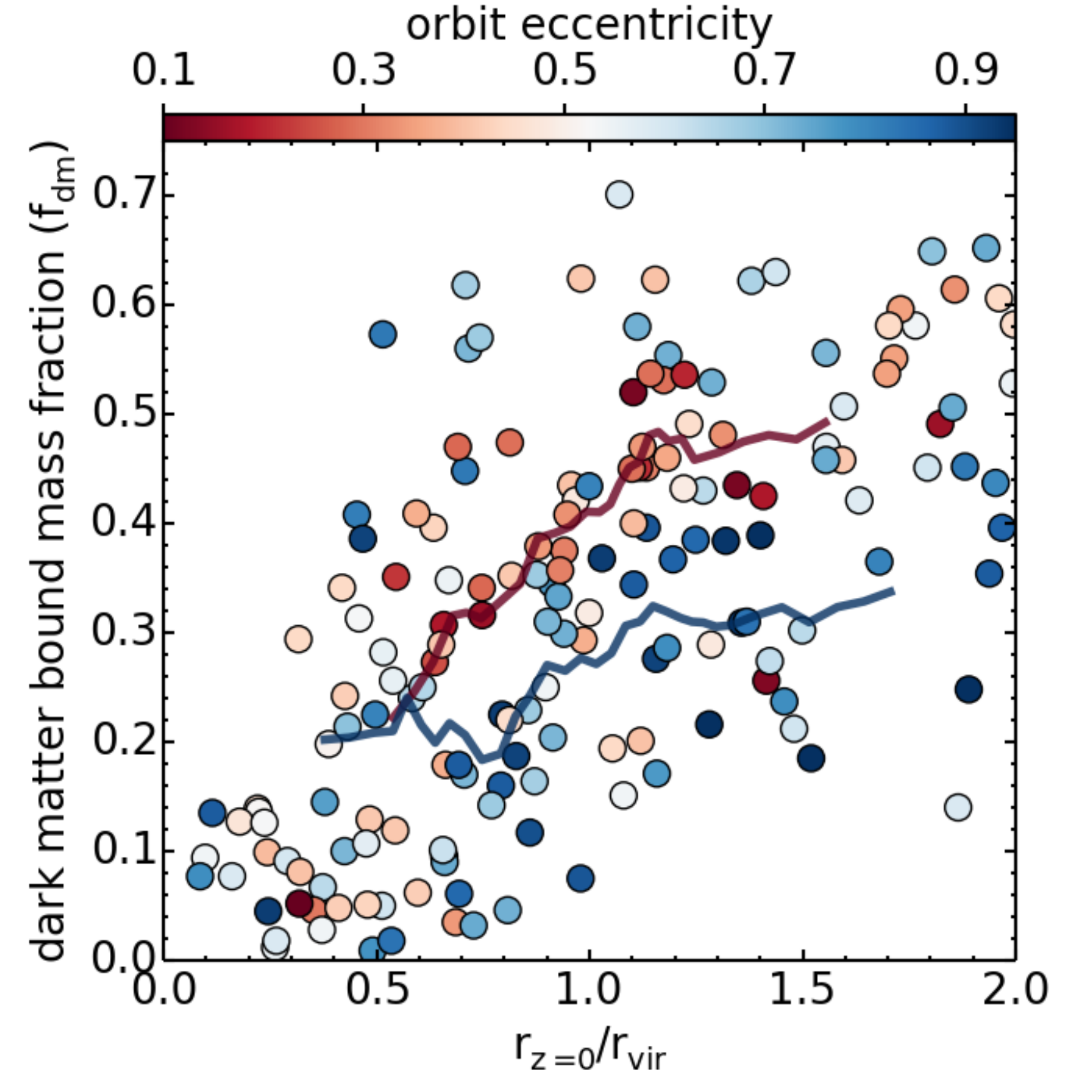}
\end{center}
\caption{Bound dark matter fraction versus (normalised) clustocentric radius from the final snapshot of this study's harassment simulations. Symbol colour indicates the eccentricity of the orbit (see colour-bar). The blue curve is a running average of data points with eccentricities $>$0.75, while the red curve is a running average of data points with eccentricities $<$0.3. There is typically an offset between the red and blue curve indicating that, at fixed radius, galaxies that have suffered the most mass loss are typically on more eccentric orbits. This occurs because eccentric orbits can bring more plunging, and so more tidally stripped, galaxies out to apocentre where they spend the majority of their time.}
\label{HeavyHarassLocation}
\end{figure}

\subsubsection{Where are the strongly harassed galaxies found now?}
In \S\ref{Howmany} we find that the strongly harassed dwarf galaxies are probably not found in substantial numbers in clusters. However where might we expect to find the small fraction that are strongly harassed? We have previously seen that strong harassment occurs primarily for orbits with small pericentres (e.g. see upper-left panel of Fig. \ref{orbitplots_masses}). However strong harassment orbits can have a broad range of eccentricities meaning their apocentres can be spread beyond the cluster core. Furthermore, galaxies with eccentric orbits spend the majority of their time at apocentre, and therefore it is inevitable that strongly harassed galaxies will be found beyond the cluster core. Consider the triangular area of `strong harassment' we defined in \S\ref{Howmany}. If we calculate the apocentre of orbits that fall on the hypotenuse of this triangle, we find a range of normalised apocentre of $r_{\rm{apo}}/r_{\rm{vir}}$=0.4-0.8. Thus we expect that strongly harassed galaxies may be found almost out to the virial radius of the cluster.

We confirm this in Fig. \ref{HeavyHarassLocation}, where we plot the final dark matter fraction $f_{\rm{dm}}$ as a function of the final clustocentric radius (normalised by virial radius) for all our harassment models. There is a clear, strong trend for decreasing bound dark matter fraction with decreasing distance to the cluster centre. However the trend is very broad, meaning that a wide range of dark matter bound fractions may be found at any particular radius. For example at the virial radius, galaxies can be found with bound fractions less than 0.1 or greater than 0.7. The most eccentric orbits (those with symbols that are more blue) are preferentially (but not uniquely) found lower in the spread. This can be more clearly seen by comparing the running averages of low (red) and high (blue) eccentricity galaxies. This occurs because galaxies spend a significant fraction of their time near apocentre. Given a group of objects at a particular radius that are all simultaneously at apocentre, those that have more eccentric orbits are able to pass closer to the cluster centre when at pericentre, and this is where the tidal field is more destructive. It is also interesting to note that galaxies with $f_{\rm{dm}}$$<$0.1 (i.e. strongly harassed galaxies) can be seen at a large range  or radii, out to near the virial radius of the cluster. This means that we cannot assume that strongly harassed galaxies are only found near the cluster centre, and thus determining if a galaxy is strongly harassed based on the clustocentic radius alone is very difficult. 

\begin{figure}
\begin{center}
\includegraphics[width=8.5cm]{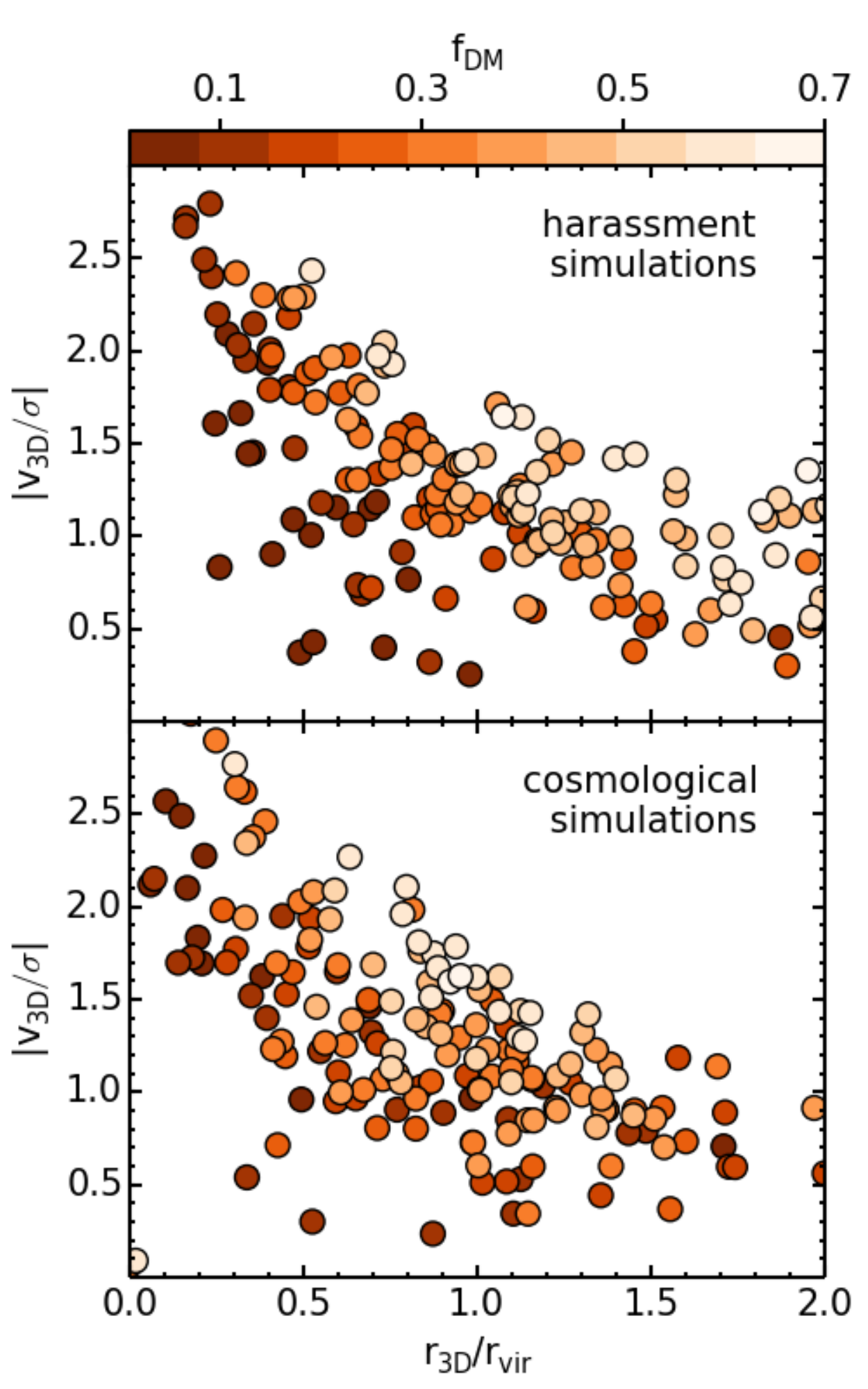}
\end{center}
\caption{Phase-space diagrams of halos in the final snapshot from our harassment model galaxies (top panel), and from the cosmological simulation of the C3 cluster of \citet{Warnick2006}  (lower panel) where we exclude first-infallers. The three-dimensional orbital velocities (y-axis) are normalised by the 3D velocity dispersion of galaxies within the cluster virial radius. The 3D clustocentric radius (x-axis) is normalised by the virial radius of the cluster. Symbol shading indicates the remaining bound dark matter fraction, with darker indicating stronger mass losses. In both the harassment simulations and cosmological simulations, the halos that suffer stronger mass-loss are systematically found at lower orbital velocity at a fixed radius.}
\label{VRCosmosims}
\end{figure}

However we can make progress if we consider the orbital velocity of the galaxy {\it{and simultaneously}} the clustocentric radius. A diagram of orbital velocity versus clustocentric radius is known as a `phase-space diagram'. In the upper panel of Fig. \ref{VRCosmosims} we show the phase-space diagram of our harassed model galaxies in the final snapshot. Galaxy orbital velocities tend to rise as the clustocentric radius becomes smaller, as expected for galaxies that fall into the potential well of the cluster and gain velocity as a result. Symbol shading indicates the remaining bound dark matter fraction, with darker symbols having suffered more mass loss. {\it{At a fixed radius, the models which lose more dark matter are systematically at lower orbital velocity.}} This is likely due to the growth in mass, and so deepening of the potential well, of the cluster with time. Thus galaxies that fall into the cluster in the past tend to have lower velocities than galaxies that infall more recently. As a result, galaxies which have spent more time being effected by the cluster also have lower velocities, and so are offset in phase-space.

In principle this suggests that a phase-space diagram analysis of galaxies might provide useful information determining statistically which galaxies are likely to have suffered more or less mass-loss from harassment. However so far we have only shown that this works in a highly idealised scenario (i.e. high speed encounters only, with no first-infallers, and a galaxy of a single halo mass and concentration). We will test if this remains valid in less ideal and more realistic scenarios in \S\ref{VRcosmosect}. The phase-space diagram also reveals that there are strongly harassed galaxies with a high velocity (i.e. at small clustocentric distance), and weakly harassed galaxies at small radius. Therefore neither a simple velocity cut, nor a simple radial cut is effective in selecting galaxies that are all strongly or weakly harassed. Instead both the velocity and radial information must be combined in a phase-space diagram to best separate between strongly and weakly harassed galaxies.

\subsubsection{Phase-space diagrams of cosmological cluster simulations}
\label{VRcosmosect}
We take the halo properties of all halos in the final snapshot of the C3 cluster from \cite{Warnick2006} (the same cluster on which our harassment model is based). We exclude first-infallers to make the plot comparable with our harassment simulations. We then measure the three-dimensional orbital velocities and clustocentric radius of each halo and plot them in the lower panel of Fig. \ref{VRCosmosims}. As in the upper panel, symbol colour denotes the fraction of dark matter that remains bound. This is calculated by dividing the final virial mass by the initial virial mass measured 7~Gyr ago. Comparing the upper and lower panel of Fig. \ref{VRCosmosims} a number of expected differences can be seen. 

In the bottom panel, the separation between galaxies that have most of their dark matter remaining and those that do not is less clear than in the upper panel. This is expected as a number of other effects are present in the cosmological simulations that do not exist in the more idealised harassment models. Halo masses vary from $\sim$$10^9$-$10^{13}$~M$_\odot$ in the cosmological simulations, and more massive halos do not necessarily suffer similar fractional mass loss as lower mass halos for the same orbit. Also in \cite{Smith2013a}, we found more concentrated halos were more robust to harassment induced mass-loss. Therefore variations in concentration and virial radius are also likely sources of scatter in the cosmological simulations. Additionally, halos that infall as part of a group in the cosmological clusters suffer mass-loss from a mixture of low and high speed encounters, whereas we deliberately excluded all orbits involving low speed tidal encounters in our harassment study.

It is remarkable, however, that despite these sources of noise {\it{halos which suffer more mass-loss are systematically shifted to lower orbital velocities at a specified radius.}} This is exciting because it means that phase-space analysis could potentially provide a highly useful tool in the future for determining if cluster galaxies have a high or low probability of having suffered strong mass-loss due to the cluster potential in real clusters. We will investigate this further in a future study, allowing for projection effects in order to provide a fair comparison with observations.


\section{Summary $\&$ Conclusions}
Using our new, cosmologically-derived harassment models, we study the influence of orbital parameters (eccentricity and pericentre distance) on the effects of harassment on early-type dwarfs in a cluster with a Virgo-like mass. We measure the fraction of stars and globular clusters that are tidally stripped. We also quantify the change in the spatial distribution of the stars and globular clusters of each dwarf galaxy model. We conduct a comprehensive study of the effects of varying the orbital parameters, consisting of 168 separate orbits of a live galaxy model within a realistic cluster potential. Our key results may be summarised as follows:
\begin{enumerate}
\item Harassment is only effective at stripping stars and globular clusters from early-type dwarfs whose orbit passes deep within the cluster core. However such orbits have a wide range of eccentricity from near-circular ($e\sim0$) to highly elongated ($e\sim1$).
\item We calculate that less than a quarter of halos in cosmological simulations, that have completed at least one pericentre passage, fall in the area of orbital parameter space where harassment is sufficiently strong to result in mass-loss of at least some stars and globular clusters (i.e. the final bound stellar and globular cluster fraction is less than one). Therefore we find that harassment is only efficient for a fraction of the orbits in our cluster. We demonstrate that this result is actually consistent with the results of \cite{Mastropietro2005}, despite appearing to be contradictory. This highlights the importance of considering the statistical probability of particular orbits when interpreting the results of harassment simulations, especially when a limited number of types of orbit are considered.
\item Due to the range of eccentricity, heavily harassed objects can be found at a large range of radii, some out to near the virial radius of the cluster. Therefore strongly harassed objects are not found exclusively near the cluster core, and so a clustocentric radius cut does not cleanly separate weakly and strongly harassed galaxies.
\item Phase-space diagrams (plots of clustocentric velocity versus clustocentric radius) are much more successful at separating weakly and strongly harassed galaxies. In our harassment simulations, we see that galaxies which suffer more mass-loss are systematically shifted to lower orbital velocities.
\item We find the same signature can be seen in phase-space diagrams of a cosmological cluster, despite the additional noise from widely ranging halo masses, halo virial radii, concentrations, and possible low speed tidal encounters. Thus we expect that a phase-space analysis of real cluster galaxies should provide information on the relative likelihood that they have suffered weak/strong mass loss due to the cluster potential.
\end{enumerate}

The stellar disks and globular cluster systems of our dwarf galaxy models are initially surrounded by a massive, and extended dark matter halo, in common with previous harassment studies, and consistent with the observed stellar to dark matter relation (\citealp{Guo2010}). Despite all our models having spent many gigayears being influenced by the cluster environment, the great majority showed no stripping or change in radial profiles of their stars and globular clusters. This is due to the fact that they are deeply embedded within the dark matter halo, and so only suffer tidal stripping when the dark matter halo is heavily tidally truncated and stripped. This only occurs for a fraction of the orbits. Therefore dwarf galaxies may not be significant contributors to the population of intra-cluster globular clusters found in the Virgo cluster (e.g. \citealp{Williams2007}; \citealp{Lee2010}; \citealp{Durrell2014}). For dwarf galaxies that are infalling into the cluster for the first time, the mass-loss can be expected to be even weaker. This could suggest that, contrary to common belief, most dwarf galaxies in similar mass clusters (e.g. Virgo) have not been significantly tidally stripped by harassment. If so, then the varied and inhomogeneous properties of early-type cluster dwarfs would likely have been set at birth, or in a group environment, rather than by the destructive tidal effects of the galaxy cluster potential.

\section*{Acknowledgements}
MF acknowledges support by FONDECYT grant 1130521. RSJ was financed through a Plaskett fellowship. Funding for this research was provided in part by the Marie Curie Actions of the European Commission (FP7-COFUND). GC acknowledges support by FONDECYT grant 3130480. RS acknowledges support from Brain Korea 21 Plus Program (21A20131500002) and the Doyak Grant(2014003730). RS also acknowledges support support from the EC through an ERC grant StG-257720, and Fondecyt (project number 3120135). T.H.P. acknowledges support by the FONDECYT Regular Project No. 1121005, Gemini-CONICYT Program No. 32100022, as well as support from the FONDAP Center for Astrophysics (15010003). MF and T.H.P acknowledge support from the BASAL Center for Astrophysics and Associated Technologies (PFB-06), Conicyt, Chile. JALA was supported by the projects AYA2010-21887-C04-04 and by the Consolider-Ingenio 2010 Programme grant CSD2006-00070. JJ thanks the ARC for financial support via DP130100388. AK is supported by the {\it Ministerio de Econom\'ia y Competitividad} (MINECO) in Spain through grant AYA2012-31101 as well as the Consolider-Ingenio 2010 Programme of the {\it Spanish Ministerio de Ciencia e Innovaci\'on} (MICINN) under grant MultiDark CSD2009-00064. He also acknowledges support from the {\it Australian Research Council} (ARC) grants DP130100117 and DP140100198. He further thanks The Lucksmiths for a little distraction. TL, RSJ, RS, and JJ gratefully acknowledge the Aspen Center Of Physics (NSF grant No. 1066293) for their great hospitality, and the valuable service they offer to visiting scientists. SKY acknowledges support from the National Research Foundation of Korea (Doyak grant 2014003730). MAB acknowledges support from the Spanish Government grant AYA2013-48226-C3-1-P and from the Severo Ochoa Excellence programme. We thank the reviewer for their thorough reading and comments.
\bibliography{bibfile}

\bsp

\label{lastpage}

\end{document}